\documentclass[11pt,titlepage]{article}
\usepackage[margin=1.2in]{geometry}
\usepackage{amsmath, amsbsy, amsthm, array, mathrsfs}
\usepackage{graphics}
\usepackage{physics}
\usepackage{amssymb,latexsym,verbatim}
\usepackage{epstopdf}
\usepackage{graphicx}
\usepackage{color}
\usepackage{dsfont, bbm}
\usepackage{relsize}
\usepackage{subcaption}
\usepackage{setspace}

\newcommand{\R}{\mathbb{R}}

\newcommand{\noin}{\noindent}
\newcommand{\bee}{\begin{eqnarray*}}
\newcommand{\ene}{\end{eqnarray*}}
\newcommand{\bec}{\begin{center}}
\newcommand{\enc}{\end{center}}
\newcommand{\be}{\begin{equation}}
\newcommand{\ee}{\end{equation}}

\newcommand{\mb}{\mathbf}
\newcommand{\bs}{\boldsymbol}
\newcommand{\tb}{\textbf}
\newcommand{\pend}{$\square$}
\newcommand{\vs}{\vskip 3mm}
\newcommand{\bi}{\begin{itemize}}
\newcommand{\ei}{\end{itemize}}

\begin{document}
\title{\LARGE On the super-efficiency and robustness of the least squares of depth-trimmed regression estimator} 
\vs
\vs
	\author{ {\sc 	Yijun Zuo and Hanwen Zuo}
		\\[2ex]
	{\small {\em Department of Statistics and Probability} and {\it Department of Computer Science
	} }\\[.5ex]
	{\small Michigan State University, East Lansing, MI 48824, USA} \\[2ex]
	{\small 
		zuo@msu.edu and zuohanwe@msu.edu}\\[6ex]
}
\date{\today}
\maketitle
\vskip 3mm
{\small
\noin	
\tb{\large Abstract}\vs
\noin
The least squares of depth-trimmed (LST) residuals regression, proposed and 
studied in Zuo and Zuo (2023), serves as a robust alternative to the classic least squares (LS) regression as well as a strong competitor to the renowned robust least trimmed squares (LTS) regression of Rousseeuw (1984). \vs
		 
The aim of this article is three-fold. (i) to reveal the super-efficiency of the LST and demonstrate it can be as efficient as (or even more efficient than) the LS in the scenarios with errors uncorrelated and mean zero and homoscedastic with finite variance and to explain this anti-Gaussian-Markov-Theorem phenomenon; (ii) to demonstrate that the LST can outperform the LTS,
the benchmark of robust regression estimator, on robustness, and the MM of Yohai (1987),
the benchmark of efficient and robust estimator, on both efficiency and robustness, 
consequently, could serve as an alternative to both;
(iii) to promote the implementation and computation of the LST regression for a broad group of statisticians in statistical practice and to demonstrate that it can be computed as fast as (or even faster than) the LTS based on a newly improved algorithm.
\bigskip

\bigskip
\par

\noindent{\bf Key words and phrase:}  depth trimming, super-efficiency,
least sum of trimmed squares of residuals, computation algorithm. 
\bigskip
\par
\noindent {\bf Running title:} least squares of trimmed residuals regression.
}
\setcounter{page}{1}		
\section{Introduction}

In the classical regression analysis, we assume that there is a relationship for a given sample $\{(\bs{x}^{\top}_i, y_i)^{\top}, i\in \{1,2, \cdots, n\}\}$: 
\be
y_i=(1,\bs{x}^{\top}_i)\bs{\beta}_0+{e}_i:=\bs{w}^{\top}_i\bs{\beta}_0+e_i~~ i\in \{1,\cdots, n\},  \label{model.eqn}
\ee
where {$e_i$ (an error term, a random variable, and is assumed to have a zero mean and unknown variance $\sigma^2$ in the classic regression theory) and $y_i$}  are in $\R^1$, ${\top}$ stands for the transpose, $\bs{\beta}_0=(\beta_{01}, \cdots, \beta_{0p})^{\top}$, the true unknown parameter, {and}~ $\bs{x_i}=(x_{i1},\cdots, x_{i(p-1)})^{\top}$ is in $\R^{p-1}$ ($p\geq 2$) and could be random.  It is seen that $\beta_{01}$ is the intercept term.  
\vs
The goal is to estimate the $\bs{\beta}_0$ based on a given sample $\mb{z}^{(n)}
:=\{(\bs{x}^{\top}_i, y_i)^{\top}, i\in\{1,\cdots, n\}\}$ from the model $y=(1,\bs{x}^{\top})\bs{\beta}_0+e$.
The difference between $y_i$ and $\bs{w^{\top}_i}{\bs{\beta}}$ is called
the ith residual, $r_i(\bs{\beta})$, for a candidate coefficient vector $\bs{\beta}$ (which is often suppressed). That is,
\be r_i:={r}_i(\bs{\beta})=y_i-\bs{w^{\top}_i}{\bs{\beta}}.\label{residual.eqn}
\ee
To estimate $\bs{\beta}_0$, the classic \emph{least squares} (LS) minimizes the sum of squares of residuals,
$$\widehat{\bs{\beta}}_{ls}:=\arg\min_{\bs{\beta}\in\R^p} \sum_{i=1}^n r^2_i. $$
\vs
The LS estimator is  very popular  in practice across a broader spectrum of disciplines due to its great computability and optimal properties when the error $e_i$ follows a normal ${N}(\bs{0},\sigma^2)$ distribution.
It, however, can behave badly when the error distribution is departed from the normal distribution,
particularly when the errors are heavy-tailed or contain outliers.
\vs
Robust alternatives to the $\widehat{\bs{\beta}}_{ls}$ are abound in the literature. The least sum of trimmed squares (LTS) of residuals regression of Rousseeuw (1984)  is the benchmark of robust regression. The idea of the LTS is straightforward, ordering the squared residuals and then trimming the larger ones and keeping at least $\lceil n/2\rceil$ squared residuals, where $\lceil ~\rceil$ is the ceiling function. Let $h$ be the number 
of the residuals kept, them the  minimizer of the sum of $h$ smallest squared residuals is called an LTS estimator:
\be
\widehat{\bs{\beta}}_{lts}:=\arg\min_{\bs{\beta}\in \R^p} \sum_{i=1}^h (r^2)_{i:n}, \label{lts.eqn}
\ee
where $(r^2)_{1:n}\leq (r^2)_{2:n}\leq \cdots, (r^2)_{n:n}$ are the ordered squared residuals and constant $h$ satisfies $\lceil n/2\rceil \leq h \leq n$.
\vs
Zuo and Zuo (2023) (ZZ23)  proposed the least sum of squares of depth-trimmed (LST) residuals estimator, a strong competitor to LTS (see Section 2) and studied
the theoretical properties of the LST and discovered that it is not only as robust as the LTS but also much more efficient. This article further reveals that LST has a super-efficiency, i.e., it can be as efficient as (or even more efficient than) the LS estimator in the scenarios with errors
uncorrelated and mean zero and homoscedastic with finite variance. This is an anti-Gaussian-Markov Theorem phenomenon. \vs
The Gauss–Markov theorem states that the LS estimator has the lowest sampling variance within the class of linear unbiased estimators, if the errors in the linear regression model are uncorrelated, have equal variances and expectation value of zero.\vs Renowned super-efficient estimators include James-Stein estimator (Stein(1956), James and Stein (1961)),  Hodge-estimator (page 109 of Vaart (1998)). In Section 4, it is revealed that LST is super-efficient relative to the LS estimator in the scenarios with errors
uncorrelated and mean zero and homoscedastic with finite variance.
With an improved algorithm, newly discovered advantages of the LST (e.g., it can run faster than LTS in addition to be more efficient than LS) will be addressed in Section 5 of this article.
\vs\noin
Section \ref{sec.2} reviews the LST. Section \ref{sec.3} investigates the differences between the LTS and the LST. Section \ref{sec.4} addresses the computation issue of the LST. Section \ref{sec.5} is devoted to
the comparison of the performance of the LST versus the LTS, the MM (Yohai 1987) and the LS via concrete examples. Concluding remarks is Section \ref{sec.6} end the article. All relevant \tb{R} code is given in 
a public depository https://github.com/left-github-4-codes/amlst. 
\section{Least squares of trimmed residuals regression} \label{sec.2}

\vs
\noin
\tb{Outlyingness (or depth) based trimming}~~
Depth (or outlyingness)-based trimming
scheme trims points that lie on the outskirts (i.e. points that are less deep, or outlying). The outlyingness  (or,
equivalently, depth) of a point x is defined to be  (strictly speaking, depth=1/(1+outlyingness) in Zuo (2003))
\be
O(x, \bs{x}^{(n)})=|x-\mbox{Med}(\bs{x}^{(n)})|/\mbox{MAD}(\bs{x}^{(n)}),  \label{outlyingness.eqn}
\ee
where $\bs{x}^{(n)}=\{x_1, \cdots, x_n\}$ is a data set in $\R^1$,  Med$(\bs{x}^{(n)})=\mbox{median}(\bs{x}^{(n)})$ is the median of the data points, and  MAD$(\bs{x}^{(n)})=\mbox{Med}(\{|x_i-\mbox{Med}(\bs{x}^{(n)})|,~ i\in \{1,2, \cdots, n\}\})$ is the median of absolute deviations to the center (median). It is readily seen that $O(x, \bs{x}^{(n)})$ is a generalized standard deviation, or equivalent to the one-dimensional projection depth (see 
 Zuo and Serfling (2000) and Zuo (2003, 2006)
 for a high dimensional version). For notion of outlyingness, cf.  Stahel (1981), Donoho (1982) and
  Donoho and Gasko (1992).
\vs
\noin
\tb{Definition of the LST}~~
For a given constant $\alpha$ (hereafter assume  $\alpha\geq 1$), $\bs{\beta}$, and $\bs{z}^{(n)}$, define a set of indexes 
\be
I(\bs{\beta})=\Big\{ i:  O(r_i, \bs{r}^{(n)}) 
\leq \alpha, ~ i\in \{1, \cdots, n\} \Big\}. \label{I-beta.eqn}
\ee
where $\bs{r}^{(n)}=\{r_1, r_2, \cdots, r_n\}$ and $r_i$ is defined in (\ref{residual.eqn}).
Namely, the set of subscripts so that the outlyingness (see (\ref{outlyingness.eqn})) (or depth) of the corresponding residuals are no greater (or less) than  $\alpha$ (or $1/(1+\alpha)$). It depends on $\mb{z}^{(n)}$ 
and $\alpha$, which are suppressed  in the notation. %
For a fixed constant $\alpha$ in the depth trimming scheme, consider the quantity
\be
Q(\bs{z}^{(n)}, \bs{\beta}, \alpha):=\sum_{i=1}^{n}r_i^2\mathds{1}\left( O(r_i, \bs{r}^{(n)}\leq \alpha\right)=\sum_{i\in I(\bs{\beta})}r_i^2,\label{objective.eqn}
\ee
where $\mathds{1}(A)$ is the indicator of $A$ (i.e., it is one if A holds and zero otherwise).
Namely, residuals with their outlyingness (or equivalently reciprocal of depth minus one) greater than $\alpha$ will be trimmed.
When there is a majority ($\geq \lfloor(n+1)/2\rfloor$) identical $r_i$s, we define MAD$(\mb{r}^{(n)})=1$ (since those $r_i$s  lie in the deepest position (or are the least outlying points)).
\vs
Minimizing $Q(\bs{z}^{(n)}, \bs{\beta}, \alpha)$ over $\bs{\beta}$, one gets the \emph{least} sum of \emph{squares} of {\it trimmed} (LST) residuals estimator,
\be
\widehat{\bs{\beta}}^n_{lst} :=\widehat{\bs{\beta}}_{lst}(\mb{z}^{(n)}, \alpha)=\arg\min_{\bs{\beta}\in \R^p}Q(\bs{z}^{(n)}, \bs{\beta}, \alpha)=\arg\min_{\bs{\beta}\in \R^p}\sum_{i\in I(\bs{\beta})}r_i^2.\label{lst.eqn}
\ee

It is seen that the LTS in (\ref{lts.eqn}) essentially employs one-sided rank based trimming scheme (w.r.t. squared residuals), whereas outlyingness (or depth) based trimming is utilized in the LST in (\ref{lst.eqn}).
Both LTS and LST employ sum of squared residuals and both trim squared residuals. Are there any differences between the two? Or rather what are the differences between the two? 
\section{Differences between the LTS and the LST}\label{sec.3}
\vs

Comparing the LST in (\ref{lst.eqn}) with the LTS  in (\ref{lts.eqn}), it is readily seen that both estimators trim residuals. However, there are at least three essential differences:
\tb{(i)}
the trimming schemes are different. The LTS employs a rank-based trimming scheme that focuses only on the relative position of points (squared residuals) with respect to (w.r.t.) others and ignores the magnitude of the point and the relative distance between points whereas the LST exactly catches these two important attributes. 
It orders data (residuals)  from a center (the median) outward and trims the points (residuals) that are far away from the center. This is known as depth-based trimming.
\tb{(ii)} Besides the trimming scheme difference, there is another difference between the LTS and the LST, that is, the order of trimming and squaring. In the LTS, squaring is first, followed by trimming whereas, in the LST, the order is reversed. \tb{(iii)} The number of trimmed squares in the LTS is fixed $(n-h)$ whereas this number in LST is not fixed, it is at most $\lfloor n/2\rfloor$, where $\lfloor x \rfloor$ is the floor function. 
	\vs
	\indent
One might still believe, albeit the three differences above, that the differences are	
negligible, especially, w.r.t. the performance of the LTS which is renowned for its best robustness.
We now examine the  performance difference w.r.t. robustness.
Performance difference between the LTS and the LST will be demonstrated
 throughout   Examples, Tables, and Figures in 
 this article.

\bec
\begin{figure}[!hb]
	\centering
	\vspace*{-7mm}
		\includegraphics[width=14cm, height=8cm]%
		{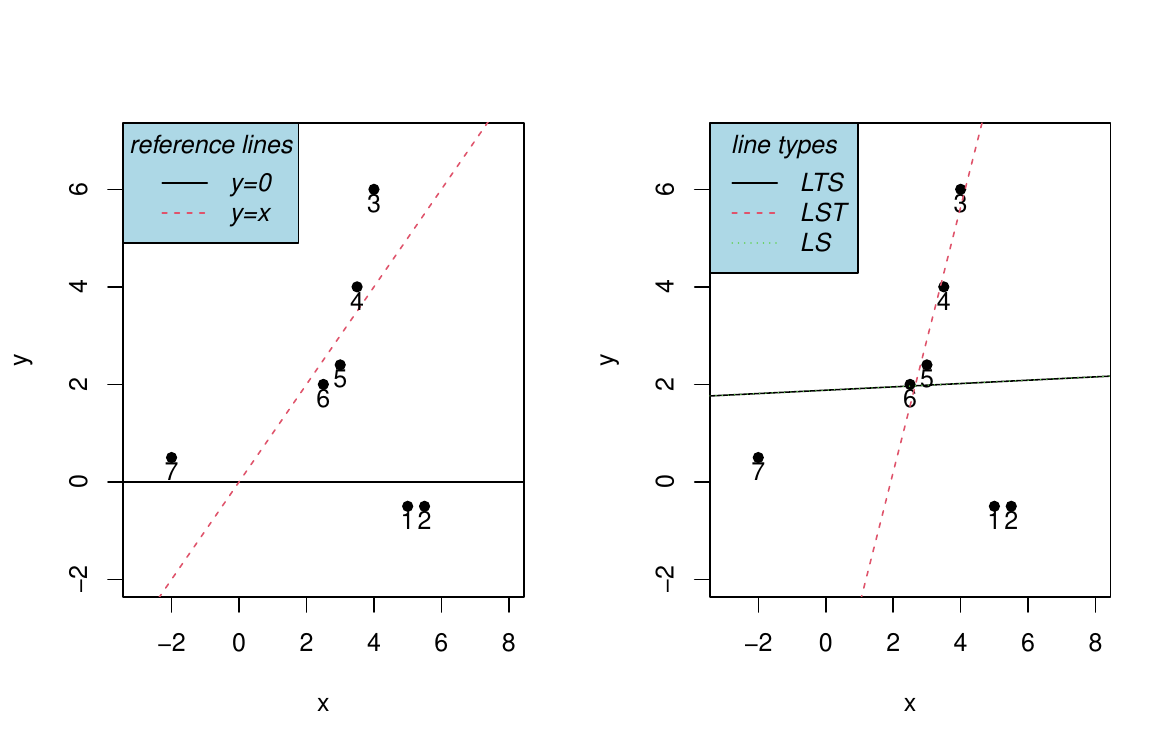}
		\caption{\footnotesize {Left: plot of seven highly correlated normal points with two outliers and two reference lines. Right: the LTS, the LST, and the LS lines w.r.t.  seven highly correlated normal points with two being contaminated. The LTS and the LS lines are identical.}}
		\label{fig.3}
		\vspace*{-5mm}
	\end{figure}
	\enc
\vspace*{-10mm}	
\tb{Robustness} In this section, let us examine the performance difference of LST and LTS on robustness. Albeit the LTS has a highest finite sample breakdown point (see, page 132 of Rousseeuw and Leroy (1987) (RL87)) but both share $50\%$ asymptotic breakdown point (see Zuo and Zuo (2023) (ZZ23)).
\vs
\noin
\tb{Example 1 (Simple linear regression)} We start with a simple illustrating toy example ($x=(5, 5.5, 4, 3.5, 3, 2.5, -2)^{\top}$, $y=(-.5, -.5, 6, 4, 2.4, 2, .5)^{\top}$), hand-calculation is feasible, the performance differences between the lines of the LTS and LST  manifest clearly. \vspace*{-5mm}
	
	\bec
	\begin{figure}[!ht]
		\centering
			\includegraphics    [width=14cm, height=8cm]%
			{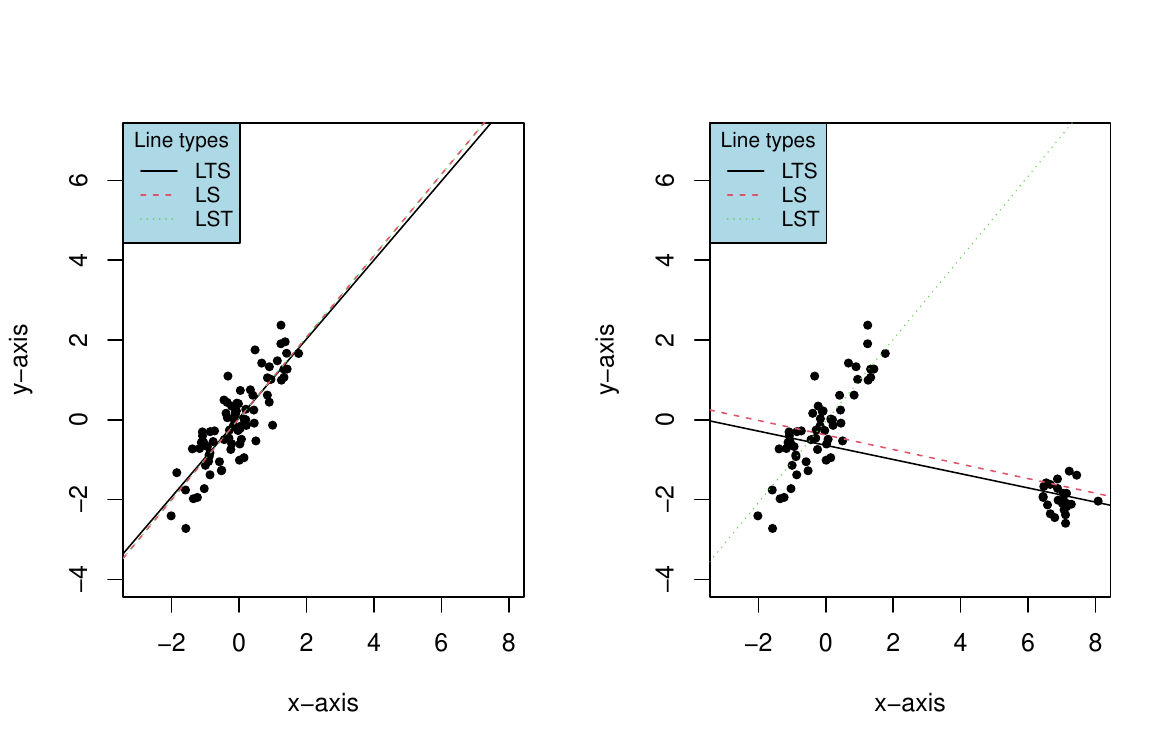}
			\caption{\footnotesize {Left: plot of three fitted lines of the LTS, the LST and the LS, to 80 highly correlated normal points. The three are almost identically catching the linear pattern. Right: the lines of the LTS and the LS are drastically changed when there are $30\%$ contamination while the line of LST resists the contamination.}}
			\label{fig.4}
			\vspace*{-4mm}
		\end{figure}
		\enc
		\noin
		\vspace*{-0mm}
		In Figure \ref{fig.3}, these seven highly correlated points with two being contaminated are plotted in the left panel along with two reference lines. In the right panel, the data are fitted by the LTS, the LST, and the LS. The lines of  LTS and LS are identical, missing the overall linear pattern which is captured by the line of LST. One shortcoming of Figure \ref{fig.3} is that the sample size is too small, the example is not practical representative and meaningful. Perhaps, the LTS and the LS will perform better when the sample size increases. This concern is met by Figure \ref{fig.4}, which again gives a negative response.
		\vs \noin
Advocators of the LTS might argue that the results in the last two Figures are not representative. 
The LTS lines should perform much better (or at least should be much more robust than the LS line when there are contamination or outliers). Is there something wrong with the calculation of the LTS lines? More in-depth analyses 
and explanations are needed.
\vs
\noin
\tb{Objective function} Both the LTS (or $\bs{\beta}_{lts}$) and the LST (or $\bs{\beta}_{lst}$)  are the minimizers of its objective functions, respectively. For the convenience of addressing the phenomenons in this and other examples, 
let us first write 
the objective function of the LTS as (see (\ref{lts.eqn}))\be \bs{O}_{lts}(\bs{\beta}):=O_{lts}(\bs{\beta}; \bs{z}^{(n)}, h)=\sum_{i=1}^h r^2_{i:n}= \sum_{i=1}^nr^2_i\mathds{1}\left(r^2_i\leq r^2_{i:h}\right),\ee  where $h$ is a fixed constant and is set to be $h=\lfloor (n+p+1)/2 \rfloor$ in \tb{R} function ltsReg; 
the one for the LST as (see (\ref{lst.eqn})) \be \bs{O}_{lst}(\bs{\beta}):=O_{lst}(\bs{\beta}; \bs{z}^{(n)}, \alpha)=\sum_{i=1}^nr^2_i\mathds{1}\left( O(r_i, \bs{r}^{(n)}\leq \alpha\right),
		\ee
where $\alpha$ is set to be three in most cases, it can be one in the pure normal error setting.
		\vs \noin
Without loss of generality, let us pick Figure \ref{fig.3} to explain the performance differences between the lines of the LTS and the LST. To fix the ideas, we will start with 
\vs\noin
\tb{Left panel of Figure \ref{fig.3}}.
One intriguing question here is, w.r.t. the two reference lines, which one does the LTS prefer in light of its objective/criterion? How does the LST?
		\vs
		\noin
Sheerly based on the trimming scheme and objective value and employing $h=\lfloor n/2 \rfloor + \lfloor (p+1)/2\rfloor=4$ (the optimal $h$, leading to highest breakdown point of the LTS, see p. 132 of RL87), the line $y=0$ (the horizontal line) has a smaller sum of four smallest squared residuals compared with that of the line $y=x$. That is, the LTS criterion prefers $y=0$ to $y=x$ among the two lines in light of the minimum sum of smallest $h$ squared residuals. This clashes with one's visual intuition. How could this happen?\vs
\noin
\tb{{Calculation}}
Let us do some direct calculations. If one looks at line $y=0$ ($\bs{\beta}_1=(0,0)^{\top}$), then it is obviously that points 3, 4, 5 have largest squared residuals and will be trimmed by the LTS criterion and it employs the sum of 4 squared residuals of points 1, 2, 6, and 7, the sum of the smallest $h(=4)$ squared residuals turns out to be  $\bs{O}_{lts}(\bs{\beta}_1)=4.75$.
		\vs
		\noin
Next, let us focus on the line $y=x$ ($\bs{\beta}_2=(0,1)^{\top}$) now.
Obviously, points 1, 2 and 7 will be trimmed and the sum of $4$ smallest squared residuals of point 3, 4, 5 and 6 will be calculated,  it turns out $\bs{O}_{lts}(\bs{\beta}_2) =4.86$, which is larger than $\bs{O}_{lts}(\bs{\beta}_1)$. Thus, in light of the objective function value,  the LTS prefers $y=0$ to $y=x$.
		\vs\noin
Up to this point there is nothing to do with the \textbf{R}  function ltsReg  used in the  calculation for the LTS line in the right panel of the figure. In ltsReg, the default value for $h$ is set to be $h=\lfloor(n+p+1)/2\rfloor$, which is 5 in this case. For details of ltsReg, refer to Rousseeuw and Van Driessen (2006). If one uses 5 smallest squared residuals,
then LTS again prefers $y=0$ to $y=x$ ($\bs{O}_{lts}(\bs{\beta}_1)=8.525$, $\bs{O}_{lts}(\bs{\beta}_2)=11.11$). On the other hand, LST  reverses this preference..
		\vs
		\noin
Detailed calculation for the LST as follows. To make the comparison fair and to employ the same number of 4 squared residuals,
w.r.t. $\bs{\beta}_1$,  points 4, 5, 6, and 7 will be kept and this leads to  $\bs{O}_{lst}(\bs{\beta}_1)=40.61$
whereas w.r.t.  $\bs{\beta}_2$, points 2, 3, 4, and 5 will be kept and this leads to  $\bs{O}_{lst}(\bs{\beta}_2)=26.01$.
Thus the LST prefers the line $y=x$ to the line $y=0$.
\vs \noin
\tb{Right panel of Figure \ref{fig.3}}.
Advocators of the LTS might hope the above can not happen in the right panel and  the LTS, searching many possible lines, should pick the line of the LST. But in reality it picks the line of the LS. This is somewhat unbelievable and unacceptable? How can this happen again?
\vs\noin
Let us first use the line of the LS as a reference line and do some direct calculations. The LTS will trim two residuals from points 2 and 3 %
and calculate the sum of five squared residuals of points 1, 4, 5, 6, and 7, obtain $O_{lts}(\bs{\beta}_{ls})=12.4062$. Now let us look at the line of LST as a reference line, then LTS will trim two residuals from points 7 and 2 and calculate the sum of five squared residuals of points
1, 3, 4, 5, and 6, obtain $O_{lts}(\bs{\beta}_{lst})=79.1716$, which is much larger than $O_{lts}(\bs{\beta}_{ls})$. That is, based on the LTS objective function or criterion, the LTS will prefer the line of LS to the line of LST (of course in the real calculation , the ltsReg searches more lines than these two).  This seems to be totally against one's visual intuition.  What is the problem? What's wrong with the LTS?
\vs \noin
\tb{The problem}
The problem is often, our eyes are cheated by Figures or {there is visual} confusion
with the residuals (the vertical distances to the line) and the Euclidean distance of points to the line.
When people inspect a Figure, they tend to interpret the closeness of points to a line in terms of {the} Euclidean distance of {the} points to the line rather than the sum of the $h$ smallest squared residuals or the selection criterion.\vs 
\noin
\tb{Tuning parameters}
That said, can one try to tune the parameter $h$ {to} get a better result? In theory, this should be feasible, but in reality, we failed to get a better result by tuning the $h$ in {ltsReg} after studying \tb{R} package robustbase. On the other hand, in this small $n$ case, one can exhaust all possible 5 points combinations ({in total} 21) and fit an LS line and calculate the sum of smallest 5 squared residuals, {to} obtain the exact solution as the ltsReg did. That is, with $h=5$, {the} current ltsReg can never result in a line {that matches the} LST.
		
\vs
\noin
In our calculation, $\alpha$ is set to be three in all cases.  In pure normal scenarios, it can  be set to be one. 
In our lstReg function, when $\alpha=1$, the total number of squared residuals employed is $\lfloor(n+1)/2\rfloor$ which is $4$ in this case.  Even with 4 residuals, the LST produces a line which looks a better fit than the one from the LTS, see the left panel {of} Figure 1 of ZZ23. (Figure \ref{fig.3} here is produced by an improved lstReg {algorithm from} AA1 in ZZ23.  The difference between the two Figures confirms the improvement).
		
		\vs
		\noin
\tb{Fair comparison}
Advocators of the LTS 
and sharp readers might argue that the comparison with the LTS is unfair unless both methods employed the same number of squared residuals (NSR). This is a legitimate concern. 
In order to get a fair comparison (with the same NSR employed),
we tuned $\alpha$  to get a total NSR employed to be 5.
It turns out if we set $\alpha=5$, then the total number meets the concern. However, the line produced is the same as the line in Figure \ref{fig.3} (which is produced with $\alpha=3$ and the total NSR employed is 4). That is, the comparison with  the LTS in Figure \ref{fig.3}  is indeed fair
(the lines in the Figure were obtained as if both procedures employed 5 smallest squared residuals). 
A similar statement holds for all other examples in this section. \hfill \pend\vs

\bec
\vspace*{-6mm}
\begin{figure}[!ht]
	\includegraphics [width=12cm, height=10cm]%
	{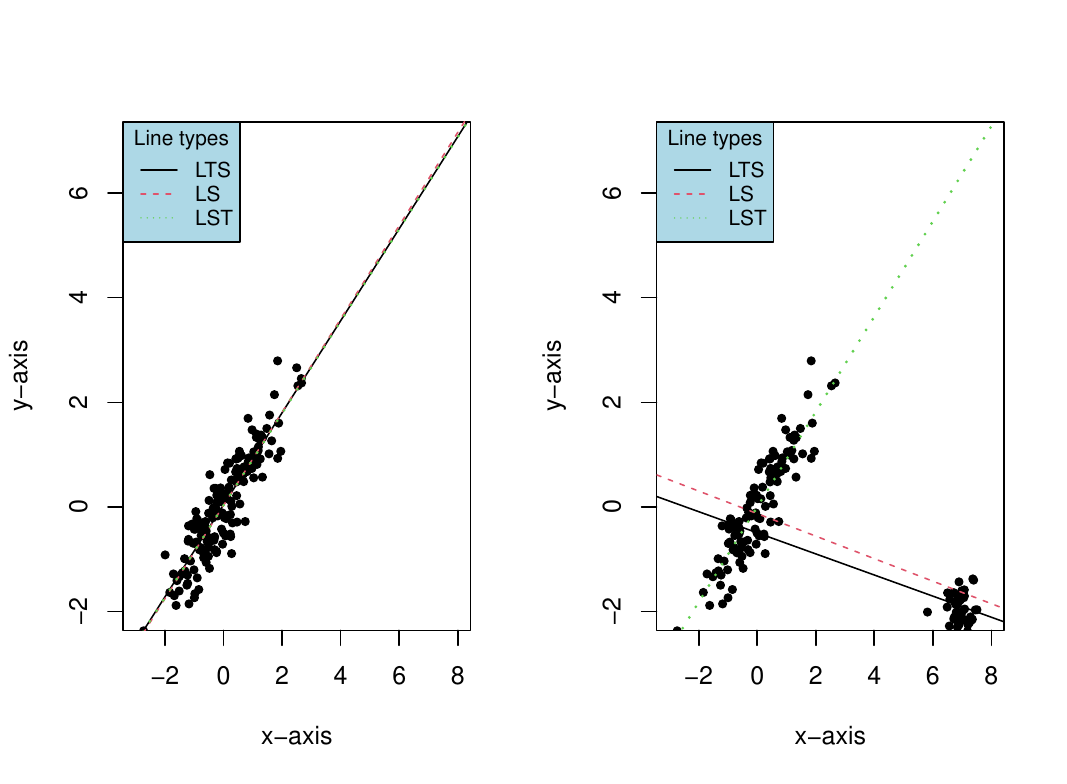}
		\vspace*{-5mm}
	\caption{\footnotesize $160$ highly correlated normal points with $30\%$ of them are contaminated by other normal points. Left: scatterplot of the uncontaminated data set and three almost identical lines. Right: the LTS, the LST, and the LS lines. Solid black is the LTS line, dotted green is the LST, and dashed red is given by the LS - parallel to LTS line in this case.}
	\label{fig.333}
	\vspace*{0mm}
\end{figure}
\enc
		\vspace*{-5mm}
		\vs \noin
\tb{Remarks}  The data set of Example 1 has been utilized in ZZ23. A similar figure to Figure 1 here also appeared in ZZ23. However, the differences include (i) here we provide detailed explanations; 	 (ii) the LST line in the right panel of Figure 1 is obtained by an improved algorithm, therefore it is quite different from the one in ZZ23.\hfill \pend
\vs

The legitimate concerns one might have include  (i) Example 1 is an isolated case; (ii) it is highly crafted, non-repeatable, non-representative; (iii) the size of data set is too small.
Next example meets the concerns since the same behavior of the LTS line repeats for the randomly generated normal
$160$ points with  $30\%$ of them being contaminated by other normal points. 
		 \vs \noin
\tb{Example 2 (Simple linear regression cont.)} 
Here again there is a contamination ($30\%$) which is way below its  upper limit (LTS theoretically allows up to $50\%$ contamination without breakdown).
		\vs \noin
In Figure \ref{fig.333} we see again that when there is 160 highly correlated normal points with $30\%$ points contaminated, the line of LTS performs 
dis-satisfactory and is ironically as sensitive to the outliers as the notorious non-robust LS line. \hfill \pend
		\vs
Two isolated examples above demonstrate the performance difference on robustness between the LTS and the LST,  perhaps they are not representative for the overall performance of the LTS. For the latter, more compelling empirical evidence is needed. We will present more various examples hereafter, especially in Section \ref{sec.6}.\vs
		
Robustness is one desired attribute of estimators, efficiency is another important desired attribute of estimators. It has been demonstrated in Zuo and Zuo (2023) that LST has an edge over LTS on the efficiency. This will be confirmed throughout this article. In the following
we address the super-efficiency of LST over the LS estimator when errors are i.i.d. $N(0, \sigma^2)$. In the robust community, MM-estimator (Yohai (1987)) is known as the benchmark of robust and efficient estimator, it will be included in our comparison hereafter.
\vs \noin

\section{Super-efficiency of LST}\label{sec.4}
LTS, LST, and MM estimators share the best $50\%$ asymptotic breakdown value while $0\%$ for the LS.  Robustness and efficiency usually do not work in tandem. For the robustness, the LTS and MM estimators have to pay a price for scarifying efficiency. LST turns out to be an exception. Gaussian-Markov Theorem states that the LS estimator has the smallest sample variance among all unbiased linear estimators when errors are uncorrelated with mean zero and homoscedastic with finite variance. To investigate the efficiency (and robustness) of LST, LTS, MM, and LS estimators, we carry a simulation 
and generate $R=1000$ samples $\{\mb{z}_i=(\bs{x}^\top_i,y_i)^\top, i\in \{1, \cdots, n\}\}$ with various $n$s and $p$s  from the  normal distribution 
with mean being a zero-vector, 
where $e_i\sim N(0,1)$ and $\bs{x}_i \sim N(\bs{0}_{p-1},  \bs{I}_{(p-1)\times (p-1)})$ are independent.
 and $y_i=(1,\bs{x}^\top_i)\bs{\beta}_0+e_i$. ($\bs{\beta}_0$ is given below).

\vs
We calculate $\mbox{EMSE} (\widehat{\bs{\beta}}):=\sum_{i=1}^R \|\widehat{\bs{\beta}}_i - \bs{\beta}_0\|^2/R$,
the empirical mean squared error (EMSE) for estimator $\widehat{\bs{\beta}}$.
Here $\widehat{\bs{\beta}}_i$ is the realization of $\widehat{\bs{\beta}}$ obtained from the ith sample with size $n$ and dimension $p$. 
Meanwhile, we also compute the sample variance of $\widehat{\bs{\beta}}$, $\mbox{SVAR}(\widehat{\bs{\beta}}):=\sum_{i=1}^R \|\widehat{\bs{\beta}}_i - \overline{\widehat{\bs{\beta}}}\|^2/(R-1)$,   where $\overline{\widehat{\bs{\beta}}}$ is the mean of all $\{\widehat{\bs{\beta}}_i, i\in\{1, 2, \cdots, R\}\}$. We obtain the finite sample relative efficiency (RE) of a procedure (denoted by P) w.r.t. the LS by the ratio of $\mbox{RE}(P):=\mbox{EMSE}(\widehat{\bs{\beta}}_{ls})/\mbox{EMSR}(\widehat{\bs{\beta}}_{P})$ (note that the sample variance ratio is another choice but sample mean squared error covers the variance and squared bias). 
At the same time, we record the total time (TT) (in seconds) consumed by different procedures for all replications.  The performance of LTS, LST, MM and LS is assessed by the four criteria: EMSE, SVAR, TT, and RE. Results are listed in Table \ref{tab.one1} and plotted in Figures \ref{fig-33}, \ref{fig-333}, \ref{fig-44}, \ref{fig-444}.
(where the $R=1000$ terms in EMSE and SVAR are plotted). In the following, two scenarios: 
pure normal and contaminated normal will be considered, respectively, for a fair comparison.
 \vs One might argue that, the comparison in the last section is unfair and the disadvantages of the performance of the LTS is sheerly due to the highly-engineered contamination scheme that is favorable to the LST.
The performance of the LTS might  be much better if there are no highly crafted  contamination scheme. Let us first look at the scenario of no contamination.
\vs
\subsection{Pure normal scenario}
\begin{figure}[!ht]
		\centering
			\includegraphics
			[width=14cm, height=8cm]{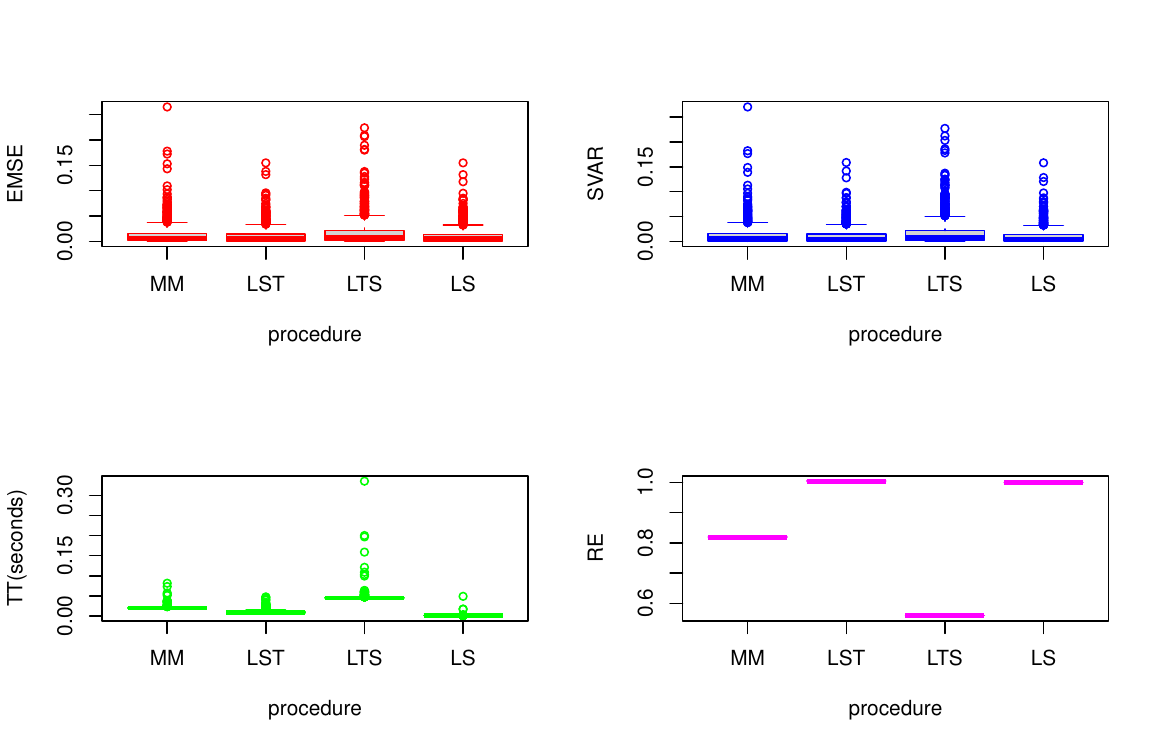}
			\caption{\footnotesize {Performance of four estimators: MM, LST, LTS, and LS w.r.t. four criteria: EMSE, SVAR, TT and RE for $p=10$, $n=100$. $\varepsilon=0\%$.
			}}
			\label{fig-33}
			\vspace*{-1mm}
	\end{figure}

\begin{figure}[!ht]
	\vspace*{-0mm}
		\centering
			\includegraphics
			[width=14cm, height=8cm]{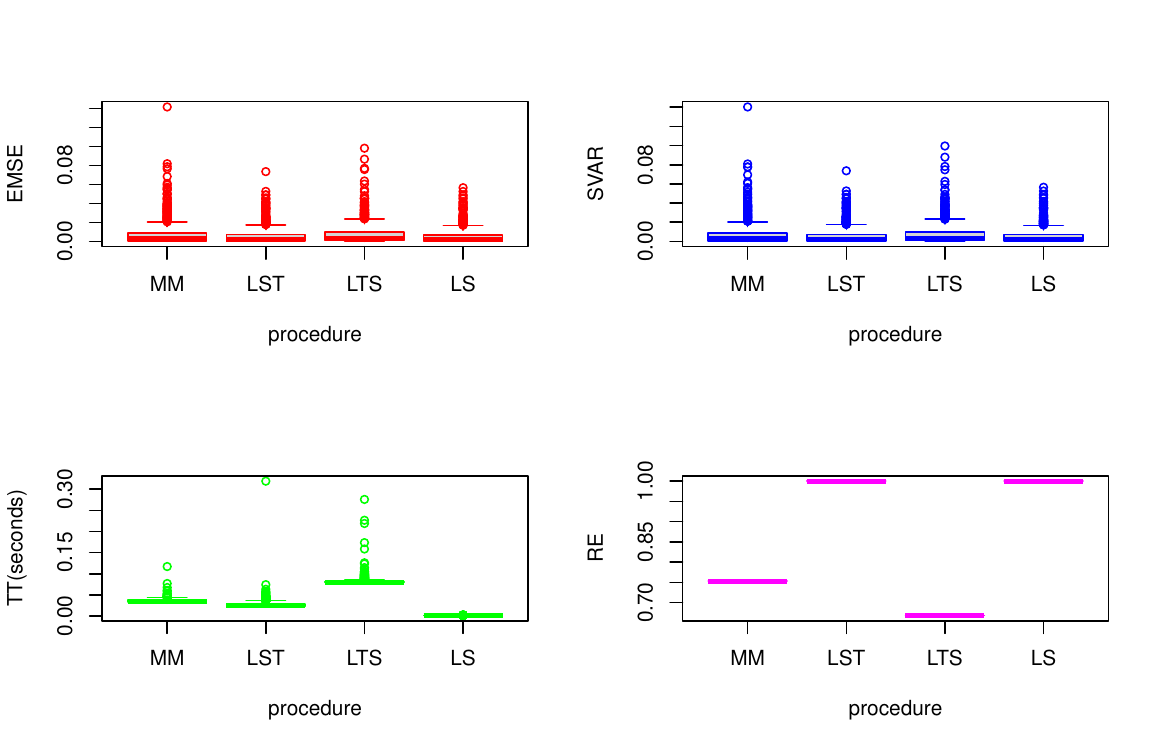}
			\caption{\footnotesize {Performance of four estimators: MM, LST, LTS, and LS w.r.t. four criteria: EMSE, SVAR, TT and RE for $p=10$, $n=200$. $\varepsilon=0\%$.
}}
			\label{fig-333}
			\vspace*{-5mm}
	\end{figure}

\begin{figure}[!ht]
		\centering
			\includegraphics
			[width=14cm, height=8cm]{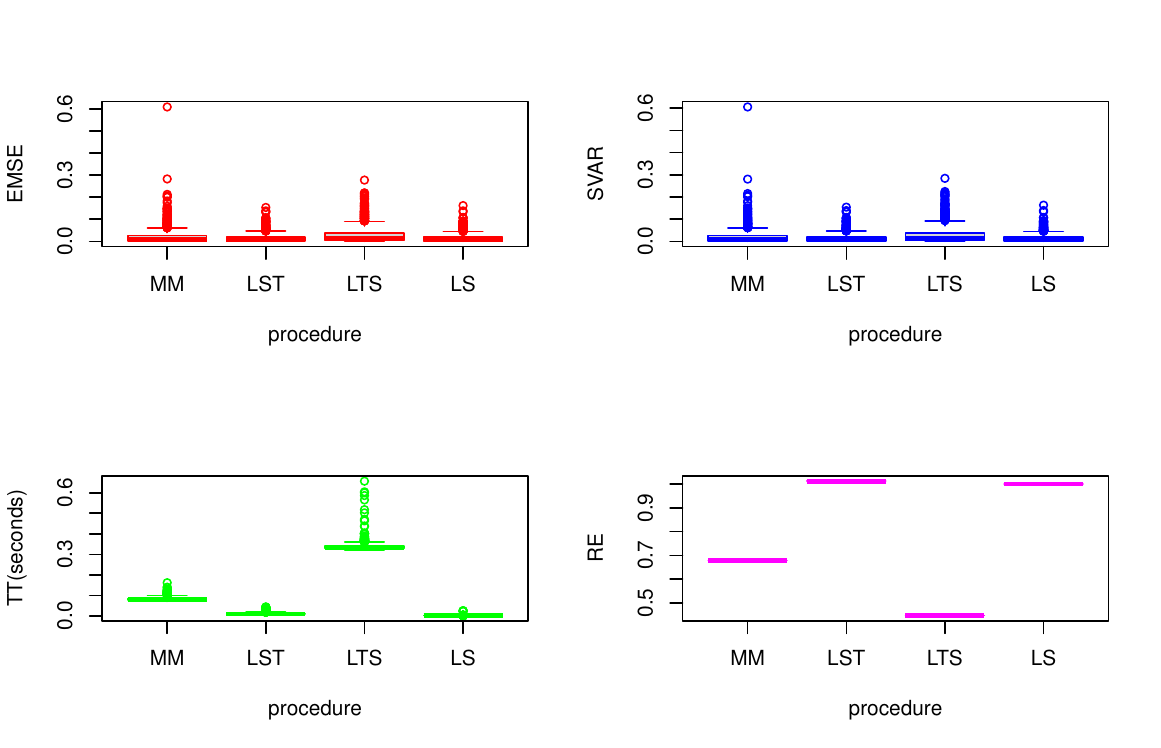}
			\caption{\footnotesize {Performance of four estimators: MM, LST, LTS, and LS w.r.t. four criteria: EMSE, SVAR, TT and RE for $p=30$, $n=100$. $\varepsilon=0\%$.
}}
			\label{fig-44}
			\vspace*{5mm}
	\end{figure}

\begin{figure}[!ht]
		\centering
			\includegraphics
			[width=14cm, height=8cm]{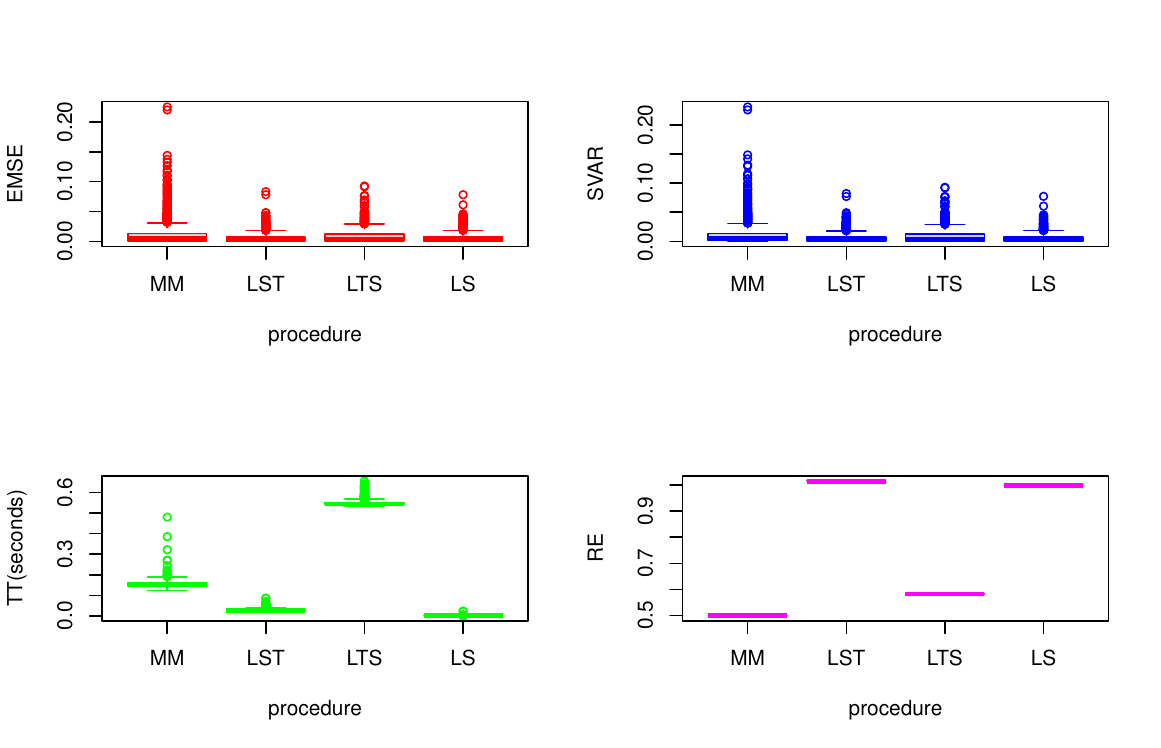}
			\caption{\footnotesize {Performance of four estimators: MM, LST, LTS, and LS w.r.t. four criteria: EMSE, SVAR, TT and RE for $p=30$, $n=200$. $\varepsilon=0\%$. 
}}
			\label{fig-444}
			\vspace*{5mm}
	\end{figure}

Here we generated $e_i\sim N(0,1)$ and $\bs{x}_i \sim N(\bs{0}_{p-1},   \bs{I}_{(p-1)\times (p-1)})$. $e_i$ and $\bs{x}_i$ are independent. Consider a sparse $\bs{\beta_0}$ such that $\bs{\beta_0}=(1, 1, 0,\ldots, 0)\in \R^p$,  and calculate $y_i=(1,\bs{x}^\top_i)\bs{\beta}_0+e_i$  obtaining $\bs{z}_i=(\bs{x}^\top_i, y_i)^\top$, $i \in\{1,2,\ldots, n\}$. We calculated EMSE, SVAR, TT, and RE for the four estimators. In light of Gaussian-Markov Theorem, no linear and unbiased estimators can have a smaller sample variance than that of LS estimator.
The requirement that the estimator be unbiased cannot be dropped, since biased estimators with lower variance  exist, see, for example, the James–Stein estimator (which also drops linearity) and ridge regression.
\vs 
\begin{table}[h]
	\centering
	\bec
	\begin{tabular}{c l c c c cccc c c}
	p&method 
	 &EMSE&SVAR&TT&RE&~~EMSE&SVAR&TT&RE\\
		\hline\\[-0.5ex]
		&  &  n=100  &$\varepsilon=0\%$ & & & n=200 &$\varepsilon=0\%$ &\\[1ex]
 &   MM&0.1376&0.1377&20.334&0.8189&0.0707&0.0706&35.502&0.7531\\
10&	LTS&0.2013&0.2014&46.216&0.5599&0.0797&0.0797&81.348&0.6683\\
&	LST&0.1123&0.1123&10.441&1.0040&0.0532&0.0532&27.463&1.0001\\
&	LS &0.1126&0.1127&1.3034&1.0000&0.0533&0.0532&1.2530&1.0000\\ [2ex]		
&  &  n=100  &$\varepsilon=0\%$ & & &n=200 &$\varepsilon=0\%$ &\\[1ex]
 &		MM &0.6379&0.6378&82.059&0.6782 &0.3524&0.3523&153.17&0.4991\\
30 &	LTS&0.9668&0.9664&338.82&0.4475 &0.3024&0.3024&550.07&0.5817\\
&		LST&0.4273&0.4272&11.142&1.0125 &0.1732&0.1732&28.089&1.0157\\
&		LS &0.4326&0.4325&1.5017&1.0000 &0.1759&0.1759&1.5638&1.0000\\[1ex]	
		\hline
	\end{tabular}
	\enc
	\caption{\footnotesize EMSE, SVAR, TT (seconds), and  RE for the MM, the LTS, the LST, and the LS based on $1000$  Gaussian samples for various $n$s and $p$s. The LST is computed by lstReg with $\alpha=3$. The LTS was computed by ltsReg. The MM was computed by lmrob.}
	\label{tab.one1}
\end{table}
Inspecting the simulation results in Table \ref{tab.one1} reveals that (i) in terms of computation speed, no one can run faster than the LS which has an overwhelming advantage. It is supposed to have the same level of advantage on efficiency. Surprisingly, LST beats LS in all cases considered. The super-efficiency of LST implied that the LST must be a biased estimator, albeit it is linear estimator (see (iii) of Theorem 2.1 of ZZ23 for linearity). (ii) MM outperforms LTS w.r.t. all four performance criteria, except in RE when $p=30$ and $n=200$ (where the renowned highly efficient MM estimator just has a $50\%$ RE, worse than $58\%$ of LTS, much worse than $102\%$ of the LST). The LST, on the other hand, has overall superiority over both MM and LTS w.r.t. all four criteria. 
\vs Numerical results in the Table \ref{tab.one1} are plotted in
Figures \ref{fig-33}, \ref{fig-333} and \ref{fig-44}, \ref{fig-444}, where 1000 terms in
the definition of EMSE and SVAR are plotted along with the total times (in seconds) consumed for each of 1000 samples by each method and overall relative efficiency (RE).
All  Figures confirm the findings (listed above (i) and (ii)) resulted from inspecting Table 1.
\vs
\subsection{Contaminated normal scenario}
\begin{figure}[!ht]
		\centering
			\includegraphics
			[width=14cm, height=8cm]{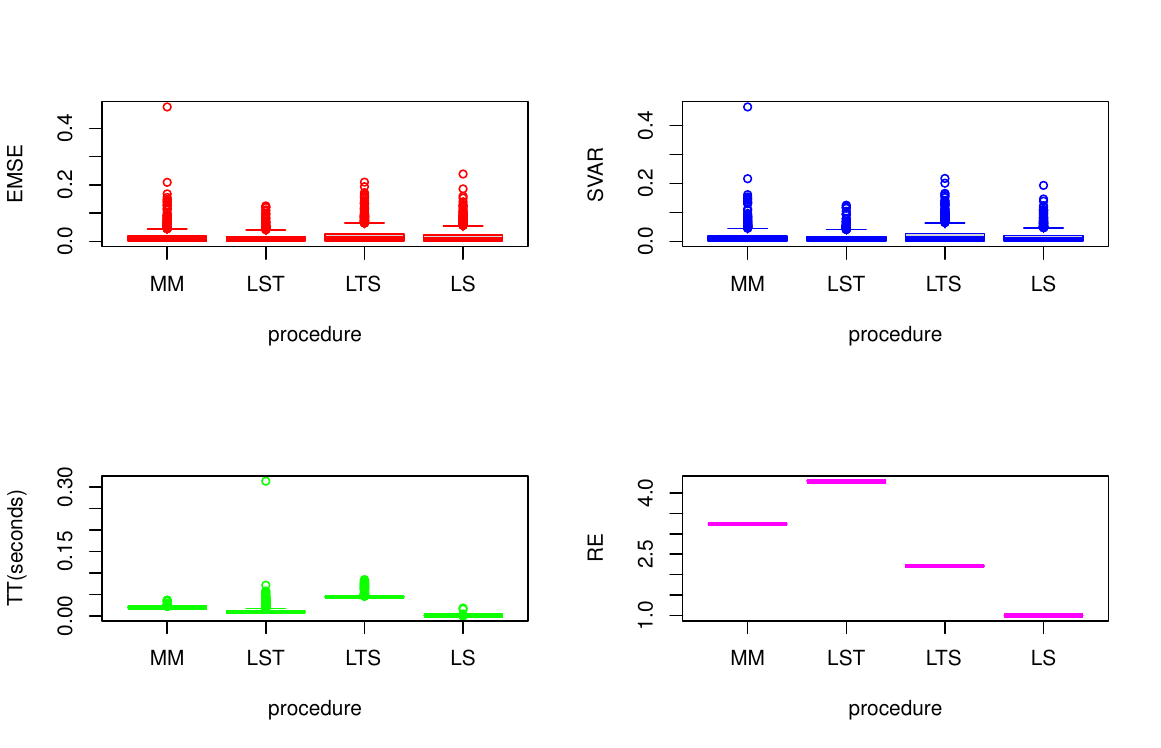}
			\caption{\footnotesize {Performance of four estimators: MM, LST, LTS, and LS w.r.t. four criteria: EMSE, SVAR, TT and RE for $p=10$, $n=100$ and $\varepsilon=20\%$ contamination.
			}}
			\label{fig.11}
			\vspace*{-5mm}
	\end{figure}

\begin{figure}[!ht]
		\centering
			\includegraphics
			[width=14cm, height=8cm]{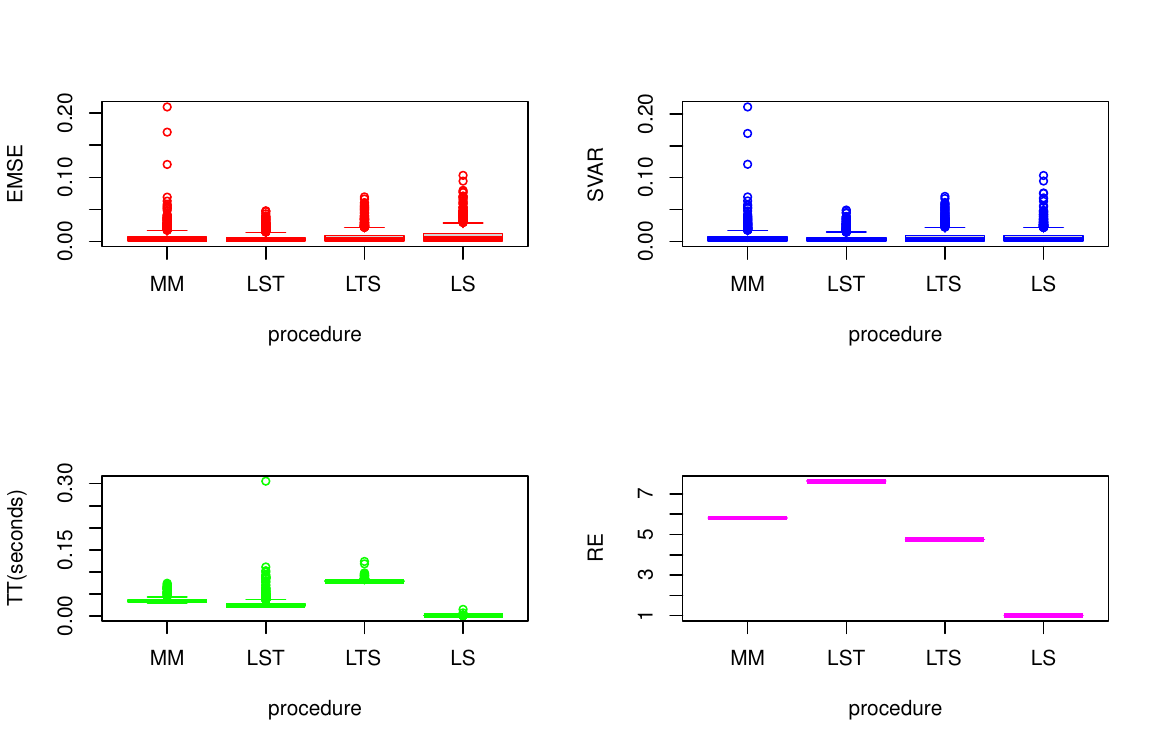}
			\caption{\footnotesize {Performance of four estimators: MM, LST, LTS, and LS w.r.t. four criteria: EMSE, SVAR, TT and RE for $p=10$, $n=200$ and $\varepsilon=20\%$ contamination. 
}}
			\label{fig.111}
			\vspace*{-5mm}
	\end{figure}

\begin{figure}[!ht]
		\centering
			\includegraphics
			[width=14cm, height=8cm]{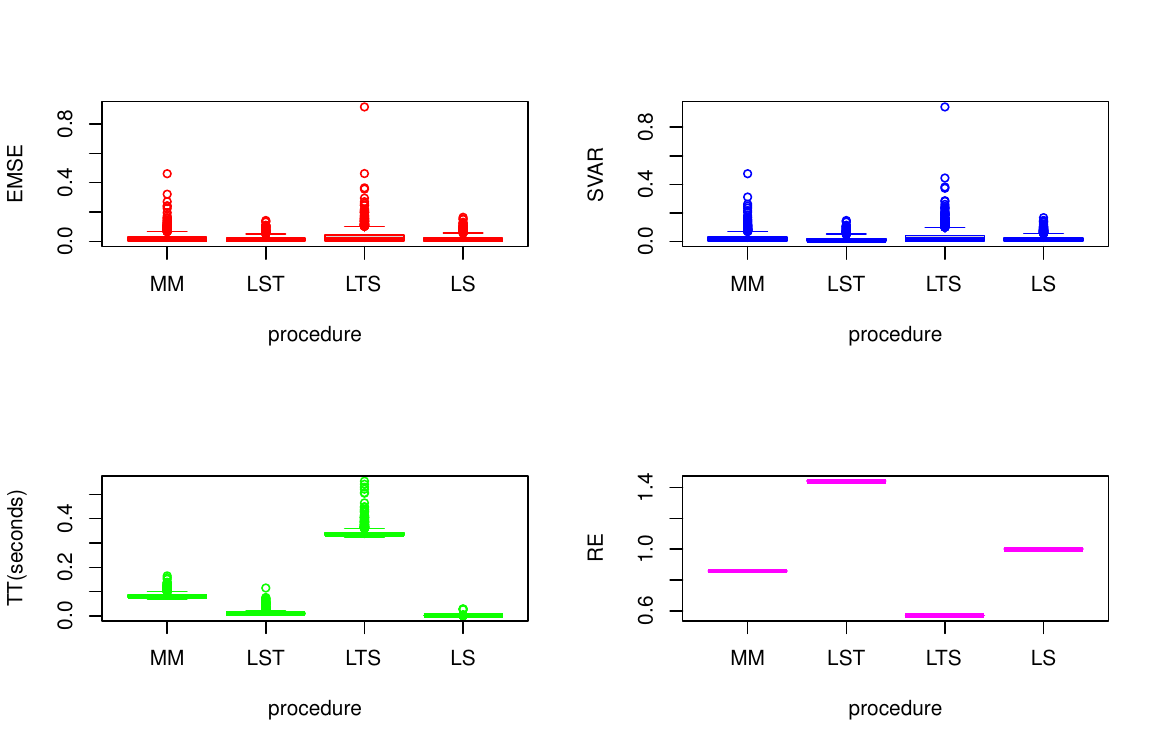}
			\caption{\footnotesize {Performance of four estimators: MM, LST, LTS, and LS w.r.t. four criteria: EMSE, SVAR, TT and RE for $p=30$, $n=100$ and $\varepsilon=20\%$ contamination.
			}}
			\label{fig.22}
			\vspace*{-5mm}
	\end{figure}

\begin{figure}[!ht]
		\centering
			\includegraphics
			[width=14cm, height=8cm]{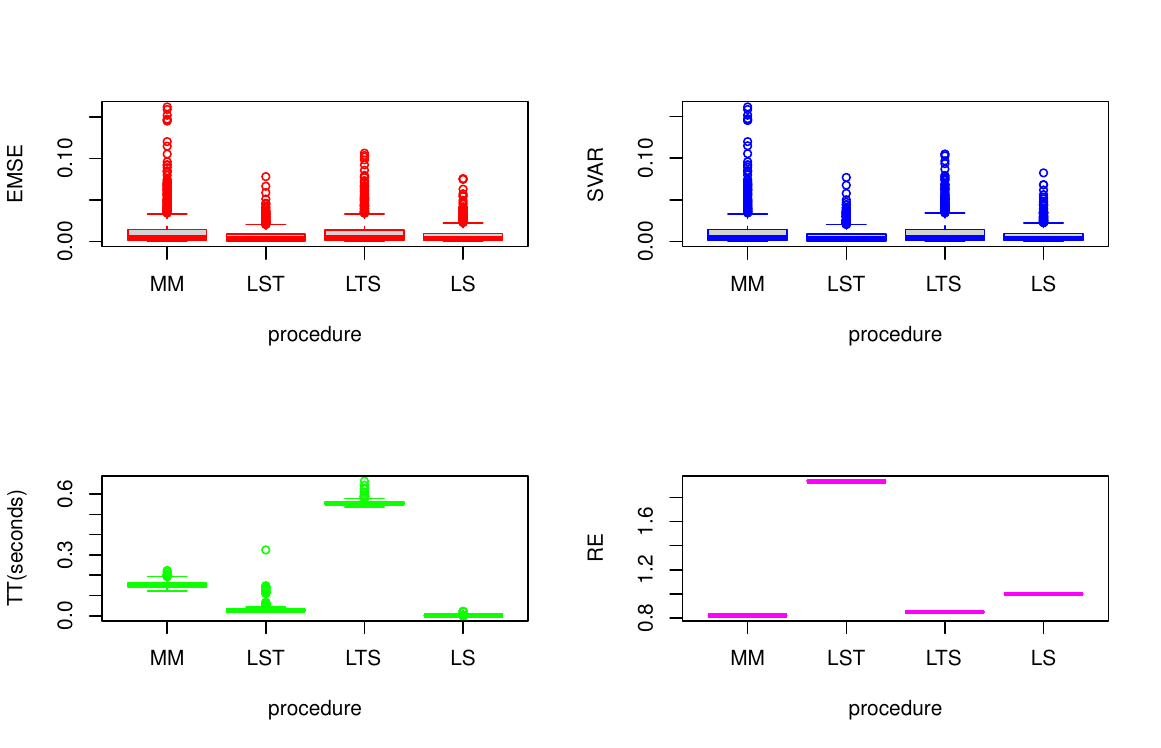}
			\caption{\footnotesize {Performance of four estimators: MM, LST, LTS, and LS w.r.t. four criteria: EMSE, SVAR, TT and RE for $p=30$, $n=200$ and $\varepsilon=20\%$ contamination. 
}}
			\label{fig.222}
			\vspace*{-5mm}
	\end{figure}

Pure normal scenario is rare in practice.
We now consider the scenario that there is an adversary contamination of sample. 
This will reveal the real picture of robustness for the four estimators. Note that theoretically speaking,  LTS, LST and MM can resist up to $50\%$ contamination, asymptotically, without breakdown.\vs We will consider the level of $\varepsilon=20\%$
contamination and generate $R=1000$ samples: 
 $e_i\sim N(0,1)$ and $\bs{x}_i \sim N(\bs{0}_{p-1}, \bs{I}_{(p-1)\times (p-1)})$. $e_i$ and $\bs{x}_i$ are independent. We consider a sparse $\bs{\beta_0}$ such that $\bs{\beta_0}=(1, 1, 0,\ldots, 0)\in \R^p$,  and calculate $y_i=(1,\bs{x}'_i)\bs{\beta}_0+e_i$,  obtain sample $\bs{z}_i=(\bs{x}^\top_i, y_i)^\top$, $i \in\{1,2,\ldots, n\}$.	Then $\varepsilon\%$ of each
sample are contaminated by $m=\lceil n\varepsilon\rceil$ points, 
we  select  $m$ points of $\{\bs{z}_i$, $i\in\{1,\cdots, n\}\}$ randomly  and replace them by $(4.5, 4.5, \ldots, 4.5, -4.5)^\top\in \R^{p}$, then apply the four methods
  to the contaminated data sets. 
Simulation results are numerically displayed in Table \ref{tab.two1} and graphically in Figures \ref{fig.11} -- 
\ref{fig.222}. \vs
\vspace*{-5mm}
	\bec
	\enc	
	\bec
	\enc	
\noin 

\vs\vs
\begin{table}[h]
	\centering
	\bec
	\begin{tabular}{c l c c c cccc c c}
	p&method 
	 &EMSE&SVAR&TT&RE&~~EMSE&SVAR&TT&RE\\
		\hline\\[-0.5ex]
		&  &  n=100  &$\varepsilon=20\%$ & & & n=200 &$\varepsilon=20\%$ &\\[1ex]
	  &  MM&0.1521&0.1519&19.917&3.2430&0.0709&0.0708&35.424&5.8113\\
	10&	LTS&0.2230&0.2226&45.979&2.2121&0.0868&0.0866&79.393&4.7464\\
	&	LST&0.1150&0.1150&12.041&4.2893&0.0540&0.0540&28.717&7.6305\\
	&	LS &0.4934&0.1381&1.2502&1.0000&0.4122&0.0658&1.2343&1.0000\\ [2ex]
&  &  n=100  &$\varepsilon=20\%$ & & &n=200 &$\varepsilon=20\%$ &\\[1ex]
 &		MM& 0.7474&0.7249&83.128&0.8595&0.4238&0.4122&153.46&0.8201\\
30 &	LTS&1.1280&1.0988&341.32&0.5695&0.4086&0.3785&556.57&0.8507\\
&		LST&0.4456&0.4455&13.087&1.4418&0.1796&0.1796&31.283&1.9353\\
&		LS &0.6424&0.4906&1.5039&1.0000&0.3476&0.1994&1.6035&1.0000\\[1ex]	
		\hline
	\end{tabular}
	\enc
	\caption{\footnotesize EMSE, SVAR, TT (seconds), and  RE for the MM, the LTS, the LST, and the LS based on $1000$  contaminated Gaussian samples for various $n$s and $p$. The LST is computed by lstReg with $\alpha=3$. The LTS was computed by ltsReg. The MM was computed by lmrob. The LS is computed by lm.}
	\label{tab.two1}
\end{table}	
Inspecting the table reveals some shocking findings (i) robust estimators MM and LTS performs worse than the non-robust LS estimator when there is $20\%$ contamination and $p=30$, they are always less than  $86\%$ efficient relative to the LS estimator. Furthermore,  if the relative efficiency calculated by the ratio of sample variance,
then both MM and LTS are always inferior to the LS estimator, they have a larger SVAR when  $\varepsilon=20\%$. Robust MM and LTS are ironically inferior to non-robust LS for the contaminated samples. 
(ii) The LST, on the other hand, has an outstanding performance.
It has the smallest EMSE and SVAR in all cases considered, it therefore has the highest relative efficiency. 
(iii) in terms of speed, no one can run faster than the LS, but LST (R-based) running faster than LTS (Fortran-based) and MM
meanwhile with a smaller EMSE and SVAR in all cases considered; (iv) LTS performs unexpectedly disappointing not only always being the slowest but also always having the worst EMSE and SVAR (except when $p=30$ and $n=200$, in that case MM is the worst performer). \vs \noin	
Advocates of MM or LTS might argue that the dis-satisfactory performance of MM or LTS is due to the highly craft contamination scheme that is favorable to the LST. The concern has its legitimacy. However, if one examines the Table \ref{tab.one1} where no contamination is involved, the superiority of the LST over MM and LTS is still outstanding.
\vs
Up to this point, we have not discussed the computing issue of the LST. How can it run faster than both MM and LTS? The latter two have the help of Fortran or Rccp. We formally address this important issue next.	
	\bec
	\enc
	\bec
	\enc	
	\vs


	\bec
	\enc
\vspace*{-10mm}

	\bec
	\enc	
	\bec
	\enc
	\bec
	\enc

\vs\vs

\section{Computation and algorithm for LST} \label{sec.5}
\tb{Exact computation}
Computation of robust estimator is usually challenging, there is no exception for the LST.
In light of Section 2 of ZZ23, one can partition the parameter space $\bs{\beta}\in \R^p$ into at most L=${n \choose k^*}$ disjoint open pieces $P_i$, $i \in \{1,2, \ldots, L\}$, $\R^p=\cup_{i=1}^L\overline{P}_i$, where $k^*=\lfloor (n+1)/2\rfloor$, $\overline{S}$ stands for the closure of set $S$. For any $\bs{\beta}$ over a single piece, the $I(\bs{\beta})$ (see (\ref{I-beta.eqn})) is identical.  
Therefore, in light of (\ref{lst.eqn}) and (\ref{objective.eqn}), the LST can actually be computed exactly (if one exhausts all pieces by picking one $\bs{\beta}$ over each piece to identify $I(\bs{\beta})$, then obtain the $\bs{\beta}_{ls}$ over the piece and  $\min$ of $O_{lst}(\widehat{\bs{\beta}}_{ls}$)) but the cost is usually not affordable. \vs \noin
\tb{Approximate computation}
We will not seek exact computation but highly approximate result by trying as many as possible pieces to obtain as good as possible approximate result. The key here is to identify each piece, or equivalently to identify a $\bs{\beta}$ from each piece. \vs
For a given sample
$\bs{z}^{(n)}:=\{(\bs{x}^\top_i, y_i)^\top, i\in \{1,2, \ldots, n\}\}$, an $\alpha\geq 1$ and a $\bs{\beta}$, one
obtains $I(\bs{\beta})$, that is, a sequence of subscripts $i_1, \ldots, i_k$ from $\{1,2, \ldots, n\}$ that constitutes of $I(\bs{\beta})$, where $k=|I(\bs{\beta})|\geq k^*$ (see Lemma 2.1 of  ZZ23), and $O_{{i_1}}<O_{{i_2}}, \ldots,
<O_{{i_k}}$, where $O_{j}:=O(r_j, r^{(n)})$ (see (\ref{outlyingness.eqn})).
Instead of directly identifying $\bs{\beta}$ for each piece, we will first identify a
$\bs{\beta}$ on the boundary of each piece $P_i$, then make small perturbation to its coordinate to obtain  two $\bs{\beta}$s from the two pieces $P_i$ and $P_j$ that share the  boundary. To find a $\bs{\beta}$ on the common boundary shared by two pieces, we notice that there must be two
indexes $s$ and $t$ such that $O_{s}=O_{t}$ and they belong to two index sets of 
$P_i$ and $P_j$, respectively. \vs
The equality of the outlyingness implies that $|r_s-\mbox{Med}(r^{(n)})|=|r_t-\mbox{Med}(r^{(n)})|$ which further implies that either (i) $r_s=r_t$ or (ii) $r_s+r_t=2\mbox{Med}(r^{n)})$. Both consequences could lead to a $\bs{\beta}$, but the (i) will be easier to maneuver. It implies that $y_s-\bs{w}^{\top}_s\bs{\beta}=y_t-\bs{w}^{\top}_t\bs{\beta}$. That is,
$y_s-y_t=(x_s-x_t)^{\top}(\beta_2, \ldots, \beta_p)^{\top}$. This implies that $\beta_1$ can be any scalar. Now if $y_s=y_t$ and $\bs{x}_s\not =\bs{x}_t$, then $\bs{\beta}=\bs{0}$
is the solution. Otherwise, if $y_s\not =y_t$ and ${x}_{s_i}\not ={x}_{t_i}$ (ith component),
then $\bs{\beta}=(\beta_1, 0,\ldots, 0, (y_{s_i}-y_{t_i})/({x}_{s_i}-{x}_{t_i}),0,\ldots, 0)^\top$ will be a solution, where, the first component could be any scalar, and the 
$(i+1)$th component is $(y_{s_i}-y_{t_i})/({x}_{s_i}-{x}_{t_i})$, all others are zeros.
Note that when $\bs{x}_i$ comes from a continuous parent distribution, then
 $\bs{x}_s\not= \bs{x}_t$ with probability one. The same is true  for $y_i$.
\vs
Summarize the discussions and ideas above, we have the following approximate algorithm.
Assume that not all points in $\{\bs{z}_i, i \in \{1, 2, \cdots, n\}\}$ are the same and $n>2$.

\vs
\noin
\tb{Step 1} \tb{Initial step: construct $2+4p$ candidate coefficient $\bs{\beta}$s}
\bi
\item[(i)] Sample two points, $\bs{x}_i$ and $\bs{x}_j$, $i\neq j, i, j \in \{1, \cdots, n\}$, until they are distinct. Assume that their $k$th components are different, that is $x_{ik}\neq x_{jk}$,  $k \in \{1, \cdots, (p-1)\}$. 
\item[(ii)] Construct two $\bs{\beta}^m$, $m\in \{0,1\}$, such that $r_i(\bs{\beta}^m)=r_j(\bs{\beta}^m)$
\[ S:=\{
\bs{\beta}^0=(0, 0, \cdots, 0, \beta_{k+1},0, \cdots, 0 )^\top;~~
\bs{\beta}^1=(1, 0, \cdots, 0, \beta_{k+1},0, \cdots, 0 )^\top\},
\]
where $\beta_{k+1}=(y_{i}-y_{j})/({x}_{ik}-{x}_{jk})$.
\item[(iii)] Perturb the $l$th component of $\bs{\beta}^m$  $m\in \{0,1\}$ with an amount $\delta$, obtained $4p$ $\bs{\beta}$s
 \[ S^0:=\{ ( \beta^0_1,\cdots,
\beta^0_{l-1}, \beta^0_{l}\pm\delta, \beta^0_{l+1},\cdots, \beta^0_p)^\top\},  
S^1:=\{ ( \beta^1_1,\cdots,
\beta^1_{l-1}, \beta^1_{l}\pm\delta, \beta^1_{l+1},\cdots, \beta^1_p)^\top\},
\]
where $l\in \{1, \cdots, p\}$ and $\delta=0.5$ or $1$. These $4p$ $\bs{\beta}$s and $\bs{\beta}^0$ and $\bs{\beta}^1$ form $(2+4p)$ $\bs{\beta}$s.
\vs
\ei
\noin
\tb{Step 2} \tb{Iteration step: compute the LSs based on sub-data sets}\vs

Set Betmat to be a $p$ by $4p+2$ matrix storing, column-wise, all $\bs{\beta}$ in step 1 above. \\
\indent
For each $\bs{\beta}$ (column) of Betmat, do
\bi
\item[(i)] Calculate $I(\bs{\beta})$. Assume that $O(r_{i_1}, r^{(n)})\leq O(r_{i_2}, r^{(n)})\leq \cdots, \leq O(r_{i_J}, r^{(n)})$ where $J=|I(\bs{\beta})|$, the cardinality of $I(\bs{\beta})$ which is identical to the set $\{i_1,i_2, \cdots, i_J\}$.
\item[(ii)] If the inequalities in (i) are not all strict, then $\{$break; go to the next $\bs{\beta}$ in Betmat$\}$\\ else 
$\{$obtain $\widehat{\bs{\beta}}_{ls}$  and sum of squared residuals (SS$_{ls})$ based on sub-data $\{\bs{z}_i, i\in I(\bs{\beta})\}$$\}$
\item[(iii)]If (SS$_{ls}<SS_{min}$), then $\{$ $\bs{\beta}_{lst}= \widehat{\bs{\beta}}_{ls}; SS_{min}=SS_{ls}$$\}$ 
\\
~~end do
\ei
\vspace*{-1mm}
~~~~~end for\\[1ex]
\noin
\tb{Step 3} \tb{Replicate steps:} Repeat steps 1 and 2 R times, R$\leq n(n-1)/2$, default value is one.

\vs
\noin
\tb{Output $\bs{\beta}_{lst}$}
\vs
\section{Illustration examples and comparison} \label{sec.6}

\subsection{Synthetic data sets}
The two examples in Section \ref{sec.3} illustrate the performance differences between the LTS and LST. Further questions remain. For example,
how does the LST perform? Or rather, is the LST robust, compared with the benchmark LTS? How efficient the LST is, compared with the renowned MM and the LS? Partial answers have been given in Section \ref{sec.4}.
Now we further answer these questions empirically by investigating the performance of the LST, comparing it with that of benchmark of robust regression, the LTS, and that of benchmark of efficient regression, the classic LS and the renowned efficient MM, through more concrete examples.
\bec
\vspace*{-6mm}
\begin{figure}[!ht]
	\includegraphics [width=12cm, height=10cm]%
	{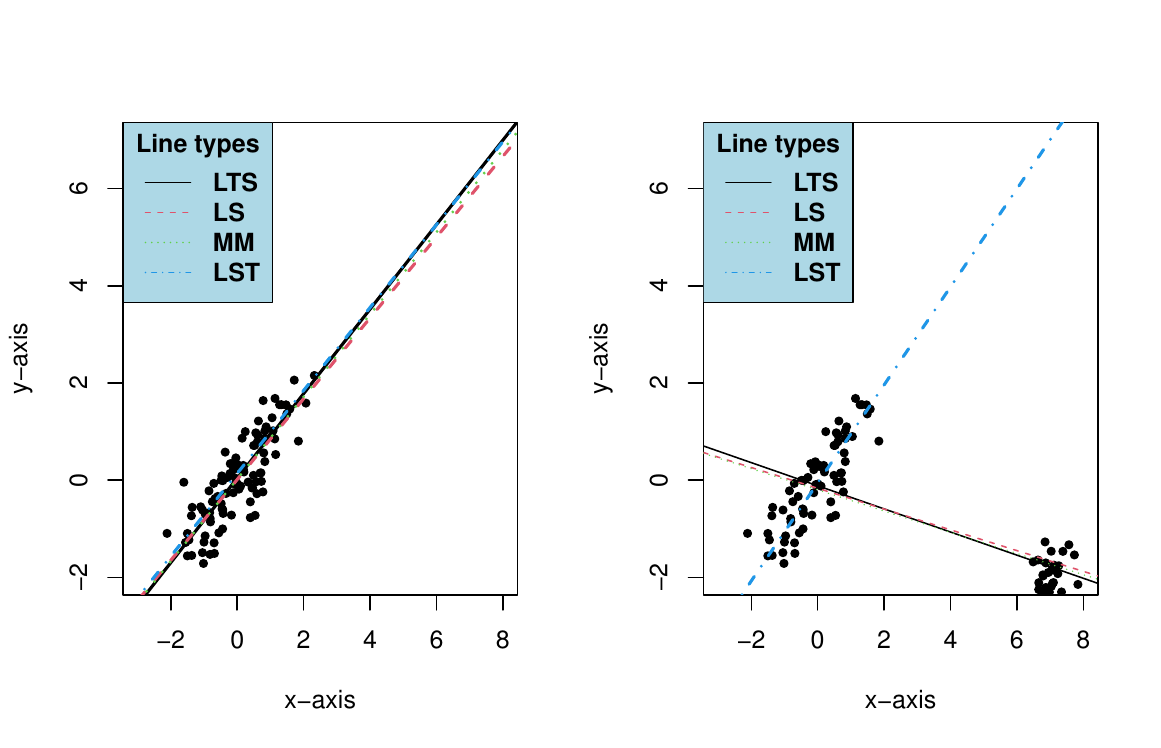}
		\vspace*{-3mm}
	\caption{\footnotesize $100$ highly correlated normal points with $30\%$ of them are contaminated by other normal points. Left: scatterplot of the uncontaminated data set and four almost identical lines. Right: the LTS, the LST, the MM, and the LS lines. Solid black is the LTS line, dashed red is the LS, the dotted green is the MM line, and the dotdash blue is given by the LST line. The first three lines are almost identical, all are drastically affected by the contamination, only the LST line resists the contamination and captures the original overall linear pattern.}
	\label{fig.two}
	\vspace*{-2mm}
\end{figure}
\enc
\bec
\vspace*{-6mm}
\begin{figure}[!ht]
	\includegraphics [width=12cm, height=10cm]%
	{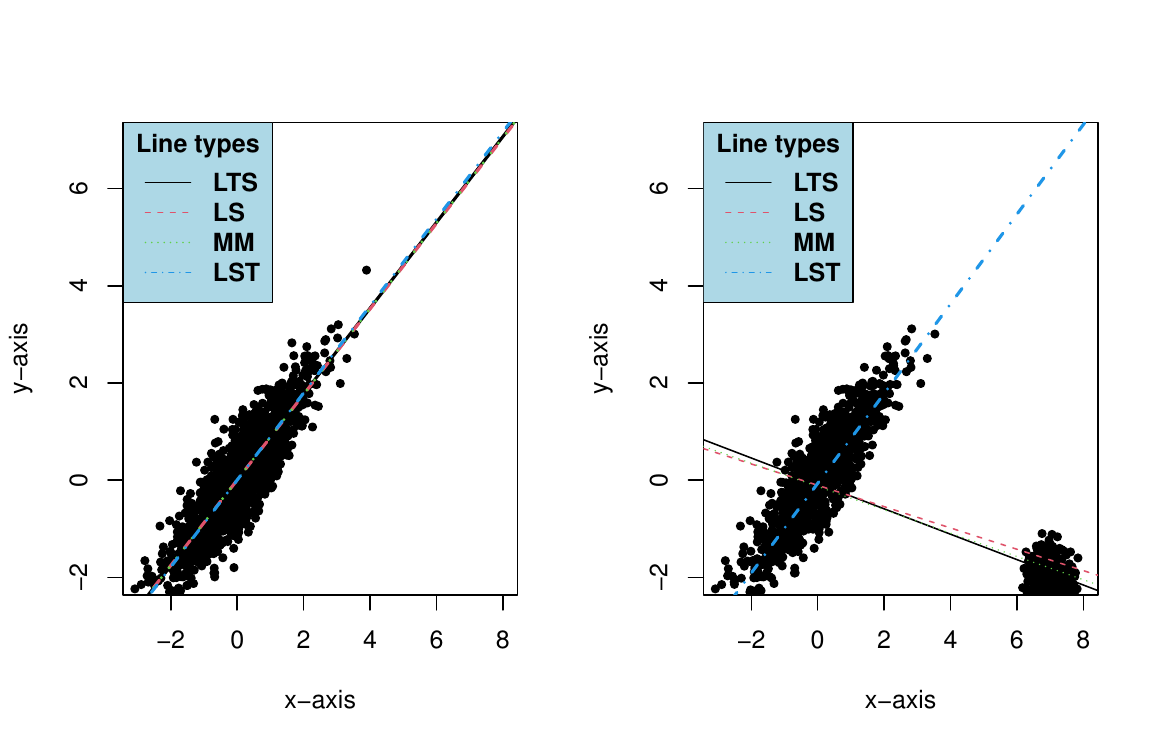}
		\vspace*{-3mm}
	\caption{\footnotesize $1600$ highly correlated normal points with $30\%$ of them are contaminated by other normal points. Left: scatterplot of the uncontaminated data set and four almost identical lines . Right: the LTS, the LST, the MM, and the LS lines. Solid black is the LTS line, dotted red is the LS, and dashed green is given by the MM. The dotdash blue line is given by the LST, the only one captures a the overall linear pattern, resisting the $30\%$ contamination.}
	\label{fig.three}
	\vspace*{-2mm}
\end{figure}
\enc

\vspace*{-5mm}
\noin
\tb{Example 3 (Illustrating bivariate data sets)}. To take the advantage of graphical illustration of data sets and plots, we again start with $p=2$, the simple linear regression.\vs

We generate $n=100$ highly correlated bivariate normal points with zero mean vector and $0.88$ as the correlation between $x$ and $y$. The scatterplot of data points along with lines fitted by four methods (LTS, LS, MM, and LST) are given in the left panel of Figure \ref{fig.two}.
The $30\%$ of points are adversarially contaminated in the right panel of the Figure. Four lines are again fitted to the contaminated data sets. 
\vs 
Inspecting the Figure \ref{fig.two} reveals that (i) all lines
catch the overall linear pattern for perfect highly correlated normal points in the left panel; (ii) in the right panel where there is a ($30\%$) contamination,
the line LTS, almost overlapping with the LS and the MM lines,  is drastically attracted by the outliers. This is not the case for the line of the LST, which resists the contamination (outliers). Note that, in theory, both LTS and MM can resist, asymptotically, up to $50\%$ contamination without breakdown, see RL87 and ZZ23. One might argue that this single example is not representative for the LTS and MM lines since the data set is still small. To meet such a legitimate concern, a similar instance is given in Figure \ref{fig.three} with sample size 
 $1600$. A similar behavior exhibits again. \hfill \pend

\vs
So far, all results displayed in Figures in the Examples 1-3 come from a single run of the methods, therefore they might not be representative. Legitimate concerns include (i) assessing the overall performance of the LTS and the MM, replications are needed;
 (ii) the contamination scheme might be highly crafted to be more favorable to the LST  than the LTS and the MM. To meet the concerns, in the next example, there will be replications and no contamination.
\vs
\noin
\tb{Example 4 (Pure Gaussian data sets)}  The least squares estimator, albeit being sensitive to outliers as shown so far, performs the best for pure Gaussian data sets. Now we investigate the efficiency of the MM, the LTS, and the LST w.r.t. the LS in the pure Gaussian 
 scenarios.\vs

For a general regression estimator $\mb{t}$,  We calculate
$\mbox{EMSE}:=\sum_{i=1}^R \|\mb{t}_i - \bs{\beta}_0\|^2/R$, the empirical mean squared error (EMSE) for $\mb{t}$. 
if $\mb{t}$ 
is  regression equivariant (see RL87 
 for definition), then one can assume (w.l.o.g.) that the true parameter $\bs{\beta}_0=\mb{0}\in \R^p$ (see RL87). 
Here $\mb{t}_i$ is the realization of $\mb{t}$ obtained from the ith sample $\bs{z}^{(n)}_i$ with $\bs{z}^{(n)}_i=\{\bs{z}_1,\ldots, \bs{z}_n\}$ and $\bs{z}_k=(\bs{x}^{\top}_k, y_k)^\top\sim N(\bs{0}, I_{p \times p})$, $i \in \{1,\ldots, R\}$, $k\in \{1, \ldots, n\}$, 
and replication number $R$ is usually  set to be $1000$. 
In addition, we also calculate  $\mbox{SVAR}(\bs{t}):=\sum_{i=1}^R \|\mb{t}_i - \bar{\bs{t}}\|^2/(R-1)$, the sample variance of $\bs{t}$,  where $\bar{\bs{t}}$ is the mean of all $\{\bs{t}_i, i\in\{1, 2, \cdots, R\}\}$, and obtain the finite sample relative efficiency (RE) of a procedure (denoted by P)  w.r.t. 
 the LS by the ratio of $\mbox{RE}(P):=\mbox{SVAR}(\widehat{\bs{\beta}}_{ls})/\mbox{SVAR}(\widehat{\bs{\beta}}_{P})$. 
Meanwhile, we record the total time (TT) (seconds) consumed by different procedures for all replications. Simulation results are listed in Table \ref{tab.one}.
\begin{table}[!h]
	\centering
	~~ Pure Gaussian data sets\\ 
	\bec
	\begin{tabular}{c l c c c cccc c}
n	&method&EMSE&SVAR&TT&RE&~~~~~EMSE&SVAR&TT&RE\\
\hline\\[-0.5ex]
&   &         &    p=10& &     &~~~  & p=15& &\\
&    MM&0.1398&0.1399&20.235&0.7999&~~0.2589&0.2589&  31.016&0.6929\\
&    LTS&0.2020&0.2020&46.312&0.5538&~~0.3294&0.3295&  86.847&0.5445\\
100& LST &0.1114&0.1114&10.559&1.0037&~~0.1782&0.1782& 10.348&1.0065\\
&    LS& 0.1118&0.1119&1.0851& 1.0000 &~~0.1793&0.1794& 1.1352&1.0000\\[1ex]
&    MM&0.0744&0.0744&35.921&0.7208&~~0.1295&   0.1294&56.113&0.6339\\
&    LTS&0.0819&0.0820&79.993&0.6543&~~0.1243&  0.1242&150.07&0.6605\\
200& LST &0.0538&0.0538&26.398&0.9960&~~0.0816& 0.0815&26.730&1.0061\\
&    LS& 0.0536&0.0536&1.0910& 1.0000 &~~0.0821&0.0820&1.1211&1.0000\\[1ex]
&   &         &    p=20& &     &~~~  & p=30& &\\
&    MM&0.2040  &0.2039&82.923&0.5426&~~0.3744&0.3744&152.65&0.4743\\
&    LTS&0.1735 &0.1735&248.43&0.6376&~~0.3024&0.3024&544.01&0.5874\\
200& LST&0.1098 &0.1097&27.628&1.0082&~~0.1768&0.1768&28.767&1.0045\\
&    LS&0.1106 &0.1106&1.2362& 1.0000&~~0.1776&0.1776&1.3284&1.0000\\[1ex]
&    MM  &0.1405&0.1405&121.73&0.5127&~~0.2854&0.1827&230.17&0.3929\\
&    LTS &0.1035&0.1035&359.70&0.6959&~~0.1708&0.1664&770.92&0.6569\\
300& LST &0.0717&0.0717&55.951&1.0042&~~0.1111&0.1073&57.288&1.0096\\
&    LS  &0.0721&0.0720&1.3075&1.0000&~~0.1121&0.1076&1.4697&1.0000\\
		\hline
    \end{tabular}
   \enc
\caption{\footnotesize EMSE, SVAR, TT (seconds), and  RE for the LTS, the LST, the MM, and the LS based on $1000$ standard Gaussian samples for various $n$s and $p$s.}
\label{tab.one}
\end{table}

\vs
Inspecting the Table \ref{tab.one} reveals some stunning findings (i) the LS is by far the fastest procedure (with 1 to 10 orders of magnitude faster), LST is the second fast runner, followed by MM, LTS is the slowest. (ii) the LS, however, is no longer the most efficient even in this pure Gaussian data points setting, the LST can be as efficient as or even more efficient than the LS (in all cases except one) (a super-efficiency phenomenon). (iii) the LST not only runs faster than LTS and MM but also possesses a smaller EMSE and SVAR than both in all cases considered. The advantages are overwhelming.
 The LST, a pure R based procedure,  could speed up by one order of magnitude if employing C++, Rcpp, or Fortran 
 just as LS, MM, or LTS does. (iv) the MM is known for its high efficiency but with a dis-satisfactory RE inferior to that of the LTS in many cases, 
  its RE could be as low as $39\%$. 
 \hfill \pend

\vs 
 Pure Gaussian points are rare in practice. The assessment above is not favorable to robust LTS and MM estimators. More realistic data points are contaminated (or mixed) Gaussian points. We now investigate the performance of the four methods against the contamination. 
\vs
\noin
\tb{Example 5 (Contaminated Gaussian data sets)}
We generate $1000$ samples $\{\mb{z}_i=(\bs{x^{\top}_i},y_i)^{\top}, i\in \{1, \cdots, n\}\}$ with various $n$s  from the  normal distribution ${N}(\bs{\mu}, \bs{\Sigma})$,
where $\bs{\mu}$ is a zero-vector in $\R^p$, and $\bs{\Sigma}$ is a $p$ by $p$ matrix with diagonal entries being $1$ and off-diagonal entries being $0.9$. Then $\varepsilon\%$ of each sample are contaminated by $m=\lceil n\varepsilon\rceil$ points.
We randomly select $m$ points from the sample $\{\bs{z}_i$, $i\in\{1,\cdots, n\}\}$ and replace them by $(7, 7, \cdots, 7, -7)^{\top}$. We apply the four methods to the contaminated samples, 
results are displayed in Table \ref{tab.two}. \vspace*{-0mm}\vs 

\begin{table}[!h]
	\centering
	~~ 
	Gaussian samples  each with $\varepsilon $ contamination rate\\
	\bec
	\begin{tabular}{l c c c c c c c c }
method  &  EMSE  & SVAR  &TT  &RE  &~~~~~EMSE  &SVAR  &TT  &RE\\
		      \hline\\[-0.5ex]
   & p=5  &  n=50  &$\varepsilon=5\%$ & &p=5 &n=50 &$\varepsilon=10\%$ &\\
MM &0.3387&0.1020&9.3336&13.424  &0.3398&0.1039&9.2514&20.444\\   
LTS&0.3877&0.1513&15.543&9.0486  &0.3728&0.1362&14.950&15.597\\
LST&0.3308&0.0935&7.6702&14.650  &0.3370&0.1006&8.9009&21.105\\
LS &1.4405&1.3692&1.2171&1.0000  &2.2985&2.1241&1.1590&1.0000\\ [1ex]
 &p=5   &n=100 & $\varepsilon=5\%$ & &  p=5   &n=100 &$\varepsilon=10\%$ & \\
MM  &0.2840&0.0480&12.809&14.960 &0.2844&0.0482&13.121&20.587\\
LTS &0.2999&0.0638&23.320&11.259 &0.2929&0.0564&23.252&17.575\\
LST &0.2821&0.0455&12.822&15.774 &0.2817&0.0452&13.895&21.932\\
LS  &0.8095&0.7183&1.0857&1.0000 &1.1564&0.9913&1.1877&1.0000\\[1ex]
\hline\\[.5ex]
		& p=10  &n=100 & $\varepsilon=20\%$& &p=10   &n=100 &$\varepsilon=30\%$& \\
		MM      &0.2439&0.1351&19.588&27.804 &4.2873&4.2486&20.475&1.0617\\
	         LTS&0.2534&0.1447&52.966&25.967 &41.480&41.357&62.586&0.1091\\
            LST &0.2409&0.1321&18.174&28.443 &0.2599&0.1516&24.039&29.757\\
	 	    LS  &3.8871&3.7566&1.0283&1.0000 &4.6634&4.5107&1.0812&1.0000\\[1ex]
         &	 p=10  &   n=200      & $\varepsilon=20\%$& &p=10&n=200 &$\varepsilon=30\%$ &\\
	      MM    &0.1687&0.0603&33.193&28.239 &0.1852&0.0771&33.059&28.136\\   
	         LTS&0.1713&0.0630&89.542&27.021 &10.232&10.222&101.32&0.2121\\
             LST&0.1674&0.0590&41.822&28.830 &0.1785&0.0700&50.045&30.966\\
		      LS&1.8341&1.7015&1.0928&1.0000 &2.3129&2.1684&1.3487&1.0000\\[1ex]
\hline\\[.5ex]
		 &p=20   &n=100  & $\varepsilon=20\%$& &p=20  & n=100 &$\varepsilon=30\%$& \\
             MM  &2.1742&2.1313&44.166&4.7494  &89.703&89.615&52.861&0.1365\\		    
		     LTS &0.7736&0.7228&268.89&14.005  &147.98&147.88&857.37&0.0827\\
            LST  &0.3711&0.3191&19.194&31.726  &0.4448&0.3930&25.438&31.134\\
		     LS  &10.218&10.123&1.1350&1.0000  &12.336&12.234&1.0876&1.0000\\[1ex]
 & p=20& n=200& $\varepsilon=20\%$ &&p=20  &n=200 &$\varepsilon=30\%$ &\\
             MM  &0.4383&0.3909&77.042&10.731  &22.281&22.206 &90.300&0.2336\\	 
		     LTS &0.1993&0.1473&373.44&28.470  &33.345&33.312&815.98&0.1557\\
             LST &0.1905&0.1385&44.868&30.286  &0.2158&0.1639&56.095&31.647\\
		     LS  &4.2869&4.1948&1.1963&1.0000  &5.2927&5.1867&1.2470&1.0000\\[1ex]
		\hline
	\end{tabular}
	\enc
	\caption{\footnotesize EMSE, SVAR, TT (seconds), and  RE for MM, LTS, LST, and LS based on $1000$ standard Gaussian samples for various $n$s and $p$ with various contamination rates.}
	\label{tab.two}
\end{table}

\vs
\begin{table}[!h]
	\centering $1000$ samples $\{ (\bs{x}^{\top}_i, y_i)^{\top}\}$ with
	$y_i=(1,\bs{x}^{\top}_i)\bs{\beta}^{\top}_0+e_i$, $\bs{x}_i \sim N(0_{p\times 1},
	 \Sigma_0)$, $e_i\sim N(0,1)$
	\bec
	\begin{tabular}{c l c c c c}
		(n,~~p) &procedure &EMSE &SVAR &TT &RE\\
		\hline\\[-0.5ex]
		&             &              & $\varepsilon=0\%$ &&\\
		&         MM   &2.5737       & 2.5748      &  230.25       &0.4033  \\ 
		&         LTS  &1.5954       & 1.5957      &  775.65       &0.6507 \\ 
		(300, 30) &LST &1.0226       & 1.0228      &  57.150       &1.0152  \\
		&    LS        &1.0381       & 1.0384      &  1.4963       &1.0000 \\[1ex]
		&             &              & $\varepsilon=5\%$ &&\\
		&         MM   &2.2144       &0.2501      &229.10      &4.6667 \\ 
		&         LTS  &2.1221       &0.1577       &820.62       &7.4021  \\ 
		(300, 30) &LST &2.0634       &0.1129        &67.096      &10.338 \\
		&    LS        &3.6486       &1.1670       &1.4914        &1.0000	\\[1ex]	           
		\hline 
	\end{tabular}
	\enc
	\caption{\footnotesize EMSE, SVAR, TT (seconds), and  RE for the MM, the LTS, the LST, and the LS based on $1000$ standard Gaussian samples, $0\%$ or $5\%$ of each sample is contaminated.}
	\label{tab.three}
\end{table}
\vs

\bec
\vspace*{-10mm}
\begin{figure}[!ht]
	\includegraphics [width=14cm, height=13cm]%
	{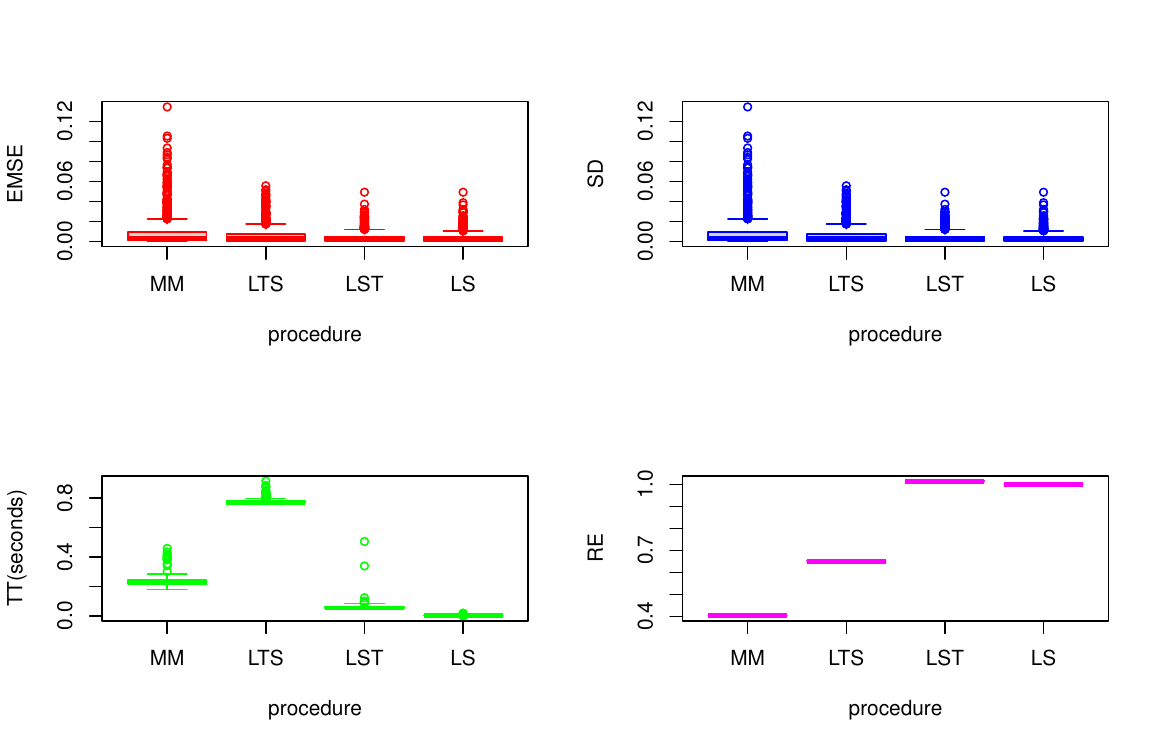}
	\caption{\footnotesize Performance of four procedures with respective $1000$ normal samples with $p=30$ and $n=300$, each sample suffers $0\%$ contamination.}
	\label{boxplot.1}
	\vspace*{-0mm}
\end{figure}
\enc

\bec
\vspace*{-14mm}
\begin{figure}[!hb]
	\includegraphics [width=14cm, height=13cm]%
	{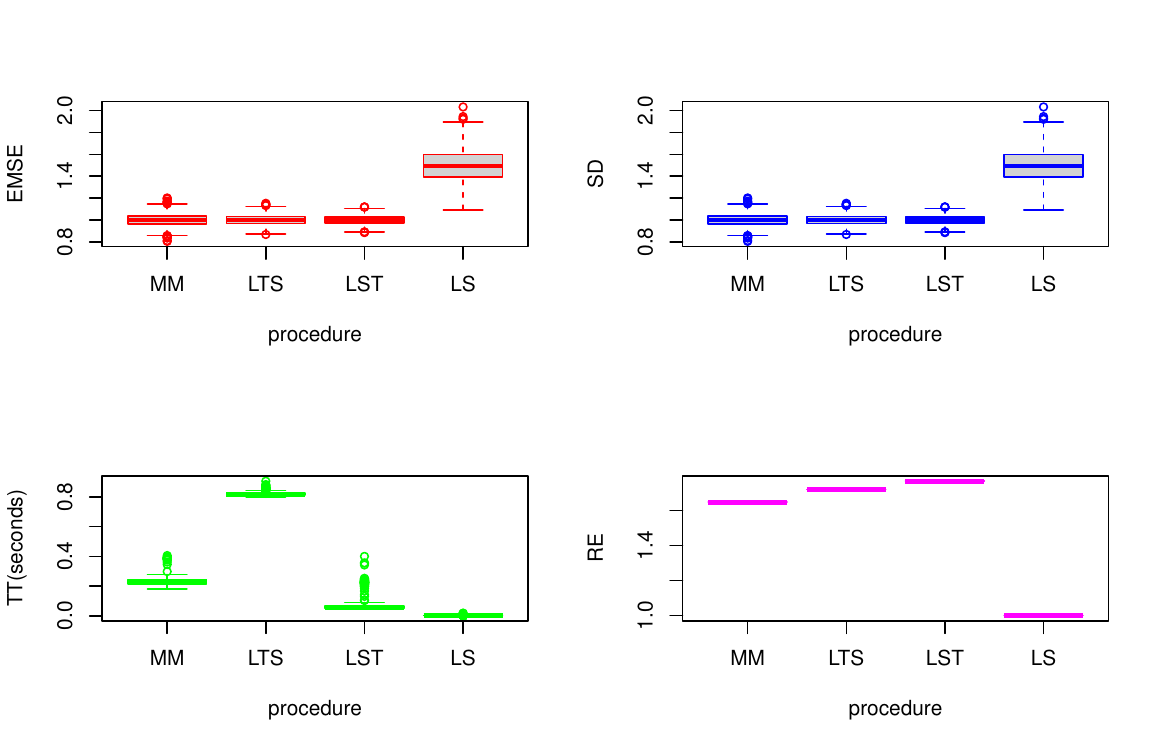}
	\caption{\footnotesize Performance of four procedures with respective $1000$ normal samples with $p=30$ and $n=300$, each sample suffers $5\%$ contamination.}
	\label{boxplot.2}
	\vspace*{-0mm}
\end{figure}
\enc
\vspace*{-7mm}

\noin
Inspecting the Table \ref{tab.two} reveals that (i) no one can run faster than the LS;
the advantage is overwhelming; (ii)compared with the LTS, the MM, and the LS, the performance of the LST is superior w.r.t. the four
criteria (except the speed); 
 (iii) With contamination, the LS is almost always the least efficient estimator with the exception when the contamination rate reaches  $30\%$ and $p=20$. In the latter case, the LTS shockingly becomes the least efficient (inconsistent-robustness), and both robust LTS and MM perform inferior to the non-robust LS estimator,
with an efficiency less than $24\%$.
(iv) The performance of LTS is frustrating when there is $30\%$ contamination and $p=20$,
the low efficiency ($8.2\%$) is consistent with its asymptotic efficiency that has been reported to be just $7\%$  in Stromberg, et al. (2000) or $8\%$ in Maronna, et al. (2006) (page 132) in the literature.  \hfill \pend

\vs
In the calculation of EMSE above, we assumed that
$\bs{\beta}_0=\bs{0}$ in light of regression equivariance of an estimator $\bs{t}$ (which holds for the four estimators). In next example, a sparse $\bs{\beta}_0$ is given.
\vs
\noin
\tb{Example 6 (\tb{Performance when $\bs{\beta}_0$ is given})}. 
We calculate $y_i$ using the formula $y_i= (1,\bs{x}^{\top}_i)\bs{\beta}^{\top}_0+e_i$, where we simulate $\bs{x}_i$ from a standard normal distribution with zero mean vector and a covariance matrix $\Sigma_0$ as the one in Example 5.
 and 
 $e_i$ from a standard normal
distribution, $i\in \{1,2, \cdots, n\}$. $e_i$ and $\bs{x}_i$ are independent. \vs

 We set $p=30$, $n=300$ and a $\bs{\beta}_0$ with its first two components being $1$ and the rest components being $0$. There is a $0\%$ or $5\%$ contamination for each of $1000$ normal samples 
  with  the contamination scheme as follows: randomly select $m=\lceil n\varepsilon\rceil$ points out of $\{\bs{z}_i$, $i\in\{1,\cdots, n\}\}$ and replace them by $(4,4, \cdots, 4, -4)^{\top} \in \R^{30}$.
We then calculate the squared deviation (SD) $(\widehat{\bs{\beta}}_i-\bs{\beta}_0)^2$ for each sample, the total time consumed by each of procedures for all $1000$ samples, and the relative efficiency (the ratio of SVAR of LS versus SVAR of a procedure). The four performance criteria, EMSE, SVAR, TT, RE, for different procedures are displayed graphically via boxplot in Figures \ref{boxplot.1} and \ref{boxplot.2} and numerically via Table \ref{tab.three}.

\vs
\indent
Inspecting the Figures and Table reveals that when there is \tb{no contamination}, (i) in terms of EMSE, LST is the best while MM is the worst and LS is the second  best. (ii)
In terms of SD (see Figure \ref{boxplot.1} 
) (or SVAR), the MM is the worst performer and LTS is the second worst performer while LST and LS are almost identically better performers (with the LST having an edge). 
(iii)
In terms of computation speed, the LS is by far the best and the LTS is clearly the worst (a fortran based function running is slower than a pure R-coded function LST). (iv) In terms of relative efficiency, LST is even more efficient (based on ratio of either EMSE, or SVAR) than LS.  When there is \tb{$5\%$ contamination},
(v) robust and efficient MM and LTS performs, ironically, inferior to the non-robust LS w.r.t. $5\%$ contaminated data sets. The superiority of the LST over the LTS, the MM, and the LS is overwhelming (the LS holds the speed advantage though).  \hfill \pend
\vs\noin
\tb{Example 7 (\tb{Performance when $\bs{\beta}_0$ is given, contin.})}.
In the above example, dimension 30 is quite high. The MM, the LTS perform not as satisfactory as one might expect. They perform might be better in lower dimensions with a high contamination rate due to their high robustness advantage. 
\vs  

\begin{figure}[!ht]
	\includegraphics [width=14cm, height=13cm]%
	{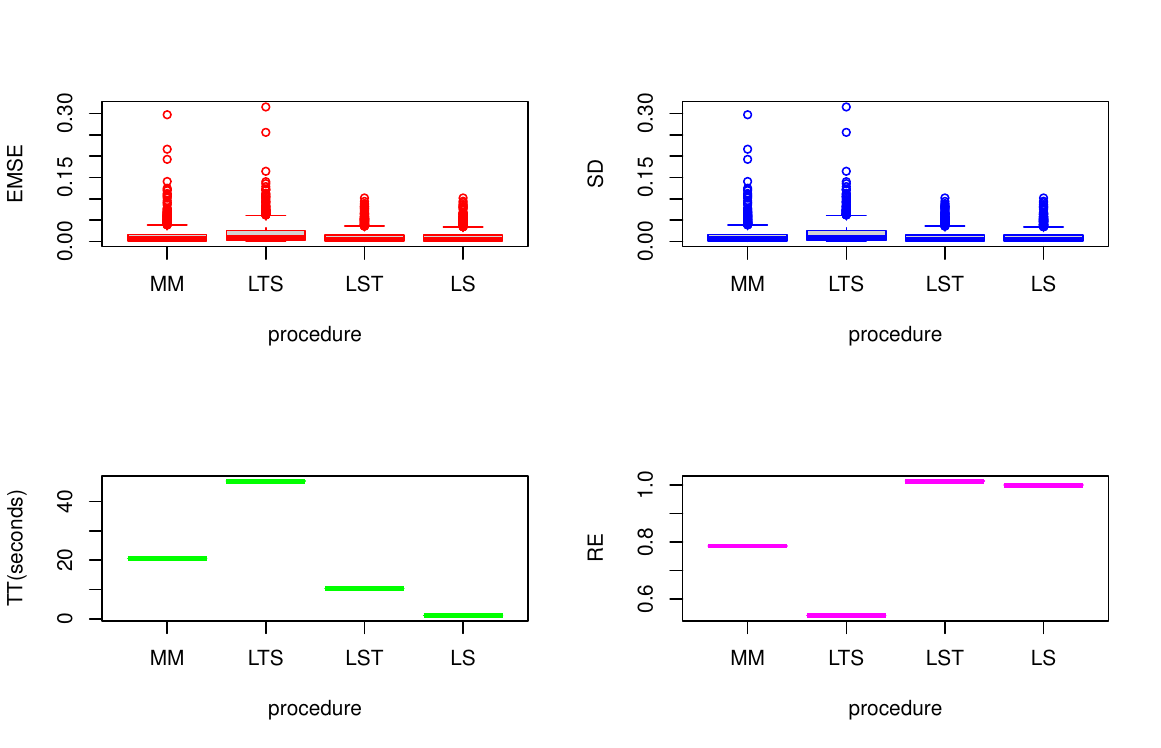}
	\caption{\footnotesize Performance of four procedures with respective $1000$ normal samples (points are highly correlated) with $p=10$ and $n=100$, each sample suffers $0\%$ contamination.}
	\label{boxplot.3}
	\vspace*{-0mm}
\end{figure}

\begin{figure}[!ht]
	\includegraphics [width=14cm, height=13cm]%
	{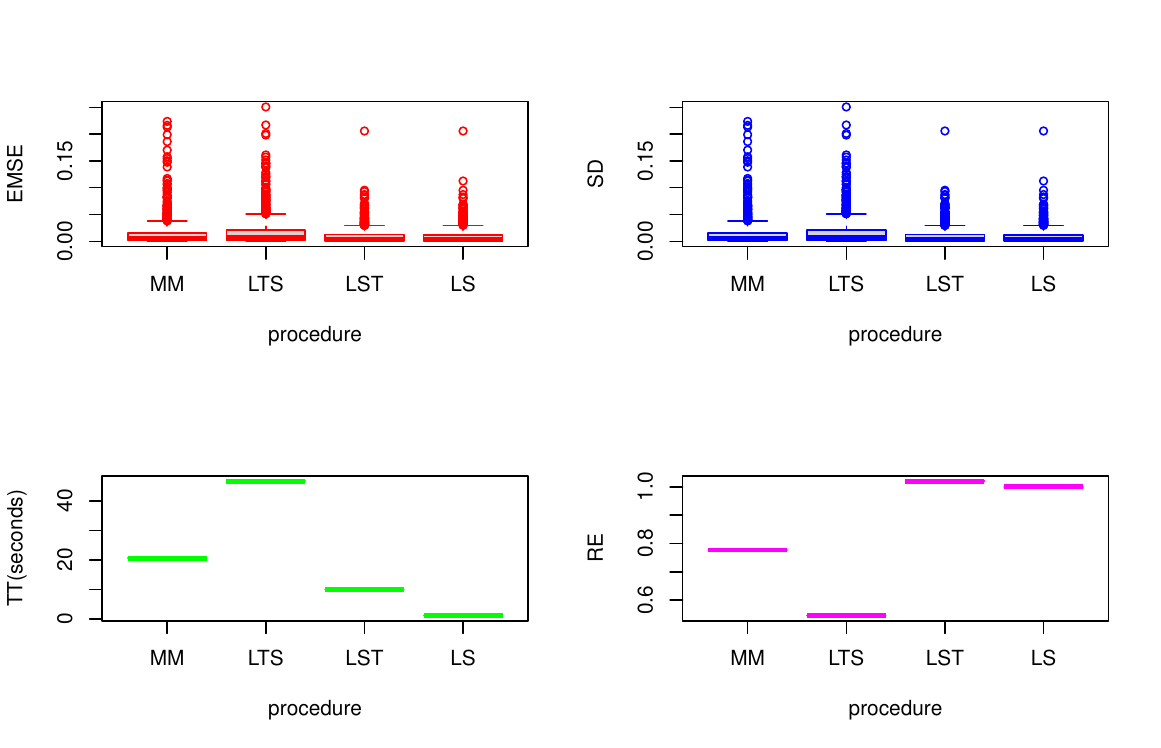}
	\caption{\footnotesize Performance of four procedures with respective $1000$ normal samples (points are highly correlated) with $p=10$ and $n=100$, each sample suffers $10\%$ contamination.}
	\label{boxplot.4}
	\vspace*{-0mm}
\end{figure}

 We set $p=10$, $n=100$ and $\bs{\beta}_0$ being a vector with its first two components being one and the rest being
zeros,  we calculate $y_i$ using the formula $y_i= (1,\bs{x}^{\top}_i)\bs{\beta}^{\top}_0+e_i$, where we simulate
 $\bs{x}_i$ from a standard normal distribution, 
that is independently from  $e_i$ which follows a standard normal distribution. \vs
There is a $0\%$ or $10\%$ contamination for each of $1000$ normal samples (generated as in Example 4) with  the contamination scheme as in Example 6. 
The  performances w.r.t.  criteria (EMSE, SD, TT and RE) are displayed 
 in Figures \ref{boxplot.3} and  \ref{boxplot.4}.\vs

Inspecting the Figures reveals that (i) even in the pure normal setting (i.i.d. errors), the LST outperforms the LS with a lower EMSE and a higher efficiency, (this stunning discovery is an anti-Gauss-Markov phenomenon); (ii) no one can run faster than the LS in both contaminated or uncontaminated settings, LST also runs faster than the LTS which is the slowest; (iii) when there is 
 $10\%$ contamination, the robust MM and LTS are unexpectedly inferior to the non-robust LS 
 (this is an inconsistent-robustness (or unstable robustness) phenomenon of the MM and the LTS). \hfill \pend\vs
 
In all above examples, data sets are synthetic or simulated. One might argue that the performance of the LTS and the MM might better for real data set.  
 \vs
\bec
\vspace*{-0mm}
\enc

\bec
\vspace*{-10mm}
\enc
\vspace*{-9mm}
\noin
\subsection{Real data sets}
\tb{Example 8 (Performance for a real data set which is known to contain outliers)}  
Here, we examine the data set of {Buxton (1920)} \cite{B20}, which has been studied repeatedly in the literature, see Hawkins and Olive (2002) \cite{HO02}, Olive (2017) \cite{O17}, Park, Kim, and Kim (2012) \cite{PKK12}, and is available at http://parker.ad.siu.edu/Olive/buxton.txt.
\vs
The Buxton data, is a 87 by 7 matrix (original row 9 was deleted due to key value missing). 
We regress y (=height in mm) on x3=head length, x4=nasal height, x5=bigonal breadth, and x6=cephalic index (two variables x1, x2, are excluded due to the missing values).
We then
fit the four different methods (mm, lst, lst, ls)
 to the data, with height as the response variable and other four variables as predictor variables. The $\hat{\beta}$ by four methods are listed in Table \ref{tab.seven}.
\vs
\begin{table}[!ht]
	\centering
	\bec
	\begin{tabular}{l c c c c c }	
		methods~&intercept &  head  &       nasal &     bigonal  &   cephalic   \\
		\hline\\[0.ex]
		MM&  1511.5503972 &  -1.1289155  &  6.5942674  & -0.6341536 &   1.2965989\\
		LTS& 1066.188018 &  -1.104774  &  6.476802   & 2.523815    &2.623706\\
		LST& 1171.385984  &  2.202086  & -2.160552 &  -1.056549  &  4.068597  \\
		LS&  1546.3737947  & -1.1288988 &   6.1133570 &  -0.5871985 &   1.1263726 
		\\[.5ex]
		\hline\\
	\end{tabular}
	\vspace*{-4mm}
	\enc
	\caption{\footnotesize Outputs of different methods based on Buxton data set.}
	\label{tab.seven}
\end{table}
Given that one does not know the true $\bs{\beta_0}$ for this real data, it is not easy to make any comments based on this single run results.
The only sensible comment one can make after
inspecting the Table is that MM performs very close to the LS (this point is solidified in below with a large replication). \vs

To distinguish the performance difference of different methods, we need replications.
One might wonder that given a fixed data set and method, the $\hat{\beta}$ given by the method
should be unique and fixed. This indeed is the case for the LST and the LS, but is not so for the MM and the LTS. The latter two do not always give a unique $\hat{\beta}$, a very serious issue in practice.
\vs

\begin{figure}[!ht]
	\includegraphics [width=14cm, height=13cm]%
	{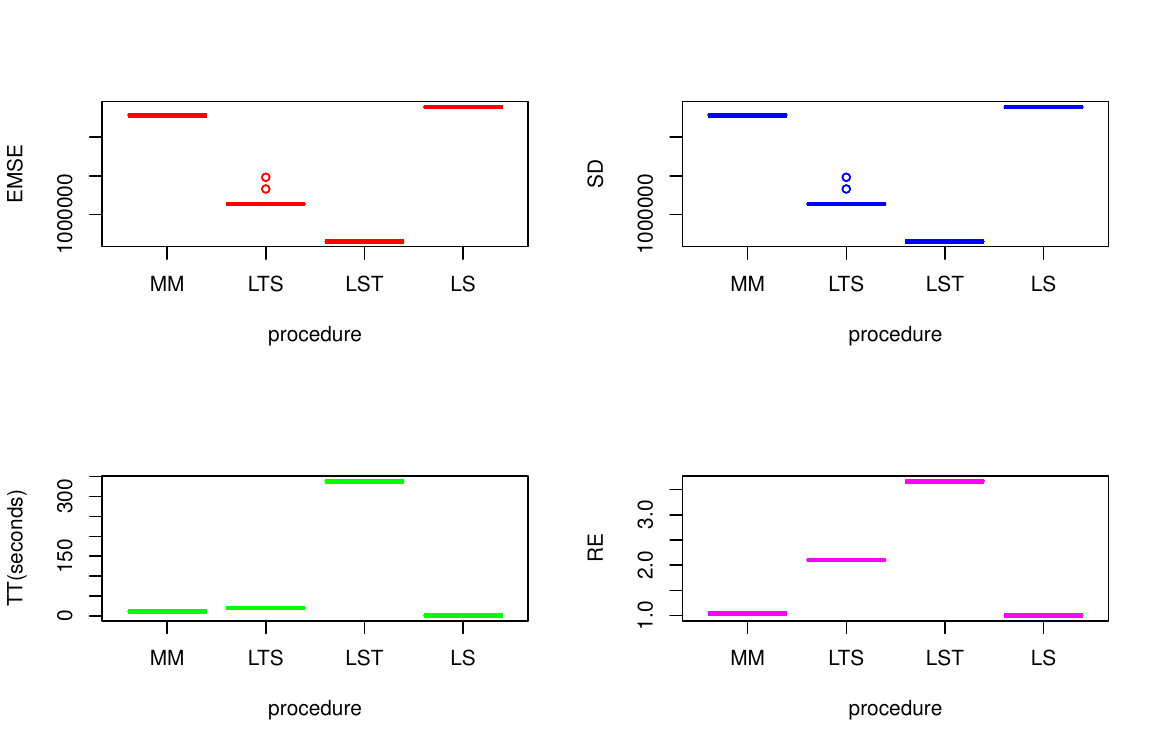}
	\caption{\footnotesize Performance of four procedures with respective to Buxton real data set}
	\label{fig-4-8}
	\vspace*{-0mm}
\end{figure}

 We repeat the calculation with calculation
R=1000 times and calculation the EMSE (with $\bs{\beta}_0$ assumed to be zero in light of regression equivalence of four method), SVAR, TT (in second), and RE (ratio of EMSE of LS
over that of the procedure). The results are listed in Table \ref{tab.eight} and displayed graphically via boxplots in Figure \ref{fig-4-8}.

\vs
\begin{table}[!h]
	\bec
	\begin{tabular}{c l c c c c}
		(n,~~p) &procedure &EMSE &SVAR &TT &RE\\
		\hline\\[-0.5ex]
		&         MM   &2284956.7      & 0.002049038      &  11.605111      &1.046546  \\ 
		&         LTS  &1137886.6       & 60.32242528      & 20.348387       &2.101538 \\ 
		(87, 5) & LST  &651824.3       & 0.000000000      &  338.66945      &3.668645  \\
		&    LS        &2391312.2      & 0.000000000     &  1.027654        &1.000000 \\[1ex]           
		\hline 
	\end{tabular}
	\enc
	\caption{\footnotesize EMSE, SVAR, TT (seconds), and  RE for the MM, the LTS, the LST, and the LS based on Buxton real data set.}
	\label{tab.eight}
\end{table}

\bec
\enc
\vspace*{-13mm}
Inspecting the Table or Figure reveals that (i) LST performs best in terms of EMSE, LS is the worst due the outliers. (ii) For a fix data set, a method should give a unique result, but this is not the case for MM and LTS, they have a non-zero sample variance. (iii) In terms of speed, LS again is the best performer and LST is the worst performer (paying a price for it lowest EMSE). (iv) In terms of relative efficiency, the LST is the best performer while the LS is worst performer, the MM is a little bit more ($105\%$) efficient than the LS. However, if one adopts the ratio of sample variance as the measure of relative efficiency, then both MM and LTS have $0\%$ relative to LS, while the LST is as efficient as the LS, has a $100\%$ RE.
\hfill \pend
\vs

\section{Concluding remarks} \label{sec.7}\vs

Is it possible to have a procedure that is as robust as the benchmark of robust regression, the LTS, meanwhile is more efficient 
 than the LTS? Is it possible to have a procedure that is as efficient as (or even more efficient than) the benchmark efficient regression, the MM, meanwhile is as robust as (or even more robust than) the MM? 
 Is it possible to have a procedure that is as efficient as (or even more efficient than) the LS in scenario of errors being uncorrelated with mean zero and homoscedastic with finite variance?  It was commonly believed that the answers to these questions were certainly negative until this article.  The latter, surprisingly, has presented promising empirically affirmative answers.
It has been asserted with some theoretical results in ZZ23 that
the least squares of depth trimmed (LST) residuals regression could serve as a robust
alternative to the least squares (LS) regression and is  a formidable competitor to the LTS regression. This article has verified the assertion empirically via concrete examples. 
It further confirmed empirically that LST is also a strong competitor to the renowned robust and efficient MM estimator.
The computation speed of the LST, albeit faster than the LTS and MM in all cases considered (except one), could be enhanced if adopting the C++, or Rcpp, or Fortran computation language as the competitors did.
\vs
Remarkable
 empirical discoveries in this article include (i) The LST is as robust as (or even more robust than) the robust benchmark LTS
while always being more efficient than the latter which can be inferior to non-robust LS when there is contamination in high dimensions; that is, the LTS has an inconsistent-robustness issue in some considered cases, besides its inefficiency issue. (ii) The LST can be as efficient as (or even more efficient than)  the benchmark LS when
the errors being uncorrelated with mean zero and homoscedastic with finite variance in high dimensions. This is a super-efficiency or anti-Gauss-Markov phenomenon. 
 (iii) The LST has an edge over MM, the benchmark of robust and efficient regression estimator, in robustness and efficiency in the cases considered in this article, the MM also suffered the inconsistent-robustness issue in some considered cases and could be inferior to non-robust LS when there is contamination in high dimensions. (iv) Both LTS and MM suffer a non-uniqueness issue, that is they could give different solutions at different time with the same data set.
 \vs
In light of all findings in this article, we conclude that the LST could serve as a robust and efficient alternative to the LS, and robust and efficient competitor to the LTS and the MM. 
\vs\vs

\noin
{\textbf {\Large Supplementary Materials}}
\vs\vs\noin
The supplementary material for this article includes the following: (A) R code for LST regression (lstReg), (B) R code for examples 1, 2, 3, 4, 5, 6, 7,  and 8. (All are downloadable at https://github.com/left-github-4-codes/amlst).
\vs\vs
\noin
	{\textbf{\Large Acknowledgments}}
\vs \vs\noin
\noin
The authors thank Prof. Wei Shao 
for 
insightful comments and stimulus discussions.
\vs\vs

\noin
{\textbf{\Large Disclosure Statement}}
\vs\vs\noin
The authors report there are no competing interests to declare.

\vs\vs
\noin
\tb{{\Large Funding}}\vs\vs\noin
This authors declare that there is no funding received for this study.

\vs\vs
\noin
\tb{{\Large ORCID}}\vs\vs\noin
Yijun Zuo: https://orcid.org/0000-0002-6111-3848
\vs\vs

{\small
\begin{thebibliography}{99}
	
	
	
	
	
	
	
	
	
\bibitem{B20} {Buxton, L.H.D.}  The Anthropology of Cyprus. \emph{J. R. Inst. Great Br. Irel.} \textbf{1920}, \emph{50}, 183--235.  
	
	
	\bibitem{D82}  Donoho, D.\ L.\ (1982), ``Breakdown properties of multivariate location estimators". PhD Qualifying
	paper, Harvard Univ.
	
	\bibitem{DG92} Donoho,\ D.\ L., and Gasko,\ M. (1992), ``Breakdown properties
	of multivariate location parameters and dispersion matrices", {\it Ann. Statist.} {\bf 20}, 1803-1827.
	


\bibitem{HO02} Hawkins, D.M.; Olive, D.J. Inconsistency of Resampling Algorithms for
High Breakdown Regression Estimators and a New Algorithm, (with discussion).
\emph{J. Am. Stat. Assoc.} \textbf{2002}, \emph{97}, 136--159.	
	
	
	
\bibitem{JS61}	
	James, W.; Stein, C. (1961), "Estimation with quadratic loss", Proc. Fourth Berkeley Symp. Math. Statist. Prob., vol. 1, pp. 361–379.
	
	
	
	
	
	
	
	
	
	\bibitem{MMY06} Maronna, R.\ A., Martin, R.\ D., and Yohai, V.\ J.(2006), `` Robust Statistics: Theory and Methods",  John Wiley \&Sons
	
	
	
    \bibitem{Netal89} Nierenberg DW, Stukel TA, Baron JA, Dain BJ, Greenberg ER(1989),  ``Determinants of plasma levels of beta-carotene and retinol",\emph{ American Journal of Epidemiology} 130:511-521.

\bibitem{O17} Olive, D.J.  \emph{Robust Multivariate Analysis}; Springer: New York, NY, USA, 2017.


\bibitem{PKK12}
Park, Y.; Kim, D.;  Kim, S. Robust Regression Using Data Partitioning
and M-Estimation. \emph{Commun. Stat. Simul. Comput.} \textbf{2012}, \emph{8},
1282--1300.
	
	\bibitem{R84} Rousseeuw,\ P.\ J. (1984), ``Least median of squares
	regression", {\it J. Amer. Statist. Assoc.} {\bf 79}, 871-880.
	
	
	
	\bibitem{RL87}Rousseeuw, P.J., and Leroy, A. (1987),  {\it Robust regression and outlier detection}. Wiley New York.
	
	\bibitem{RS98}Rousseeuw, P.\ J., Struyf, A. (1998), ``Computing location depth and regression depth in higher dimensions", {\it Statistics and Computing}, 8:193-203.
	
	\bibitem{RVD06}	Rousseeuw, P. J. and Van Driessen, K. (2006), ``Computing LTS Regression for Large Data Sets", {\it Data Mining and Knowledge Discovery} 12, 29-45.
	
	
	\bibitem{S81} Stahel, W.\ A.\ (1981), Robuste Schatzungen: Infinitesimale Optimalitiit und Schiitzungen von
	Kovarianzmatrizen. Ph.D. dissertation, ETH, Zurich.
	
	\bibitem{S56}  Stein, C. (1956), "Inadmissibility of the usual estimator for the mean of a multivariate distribution", Proc. Third Berkeley Symp. Math. Statist. Prob., vol. 1, pp. 197–206.
	\bibitem{SHH00}Stromberg, A. J., Hawkins, D. M., and H\"{o}ssjer, O. (2000), ``The Least Trimmed Differences
	Regression Estimator and Alternatives", \textit{J. Amer. Statist. Assoc.}, 95, 853-864.
	
	
	\bibitem{VDV98}Van Der Vaart, A. W.  (1998), \emph{Asymptotic Statistics}, Cambridge University Press.
	
	
	
	
	
	
	
	
	
	
	
	\bibitem{Z03} Zuo, Y. (2003) ``Projection-based depth functions and associated medians'',
	\emph{Ann. Statist.}, 31, 1460-1490.
	
	
	
	
	\bibitem{Z06} Zuo, Y. (2006), ``Multi-dimensional trimming based on projection depth", \emph{Ann. Statist.}, 34(5), 2211-2251.
	
	\bibitem{Z21a} Zuo, Y. (2021), ``On general notions of depth for regression”
	\emph{Statistical Science} 2021, Vol. 36, No. 1, 142–157, arXiv:1805.02046.
	
	
	\bibitem{Z23} Zuo, Y. (2023), ``Non-asymptotic robustness analysis of regression depth median", {\it Journal of Multivariate Analysis}, https://doi.org/10.1016/j.jmva.2023.105247, arXiv: 2009.00646
	\bibitem{ZS00} Zuo, Y., Serfling, R., (2000), ``General notions of statistical depth function", {\it Ann. Statist.}, 28, 461-482.
	
	\bibitem{ZZ23} Zuo, Y. and Zuo, H. (2023), ``Least sum of squares of trimmed residuals regression", {\it Electronic Journal of Statistics}, Vol. 17, No. 2, 2416-2446.
	arXiv:2202.10329.
\end{thebibliography}

}
\end{document}